\documentclass[a4paper,fleqn,final]{cas-sc}

\usepackage{float}
\usepackage{caption}
\usepackage[section]{placeins}

\ExplSyntaxOn
\cs_gset:Npn \__first_footerline:
  { \group_begin: \small \sffamily \__short_authors: \group_end: }
\ExplSyntaxOff

\usepackage[square,numbers,sort&compress]{natbib}

\def\tsc#1{\csdef{#1}{\textsc{\lowercase{#1}}\xspace}}
\tsc{WGM}
\tsc{QE}
\tsc{EP}
\tsc{PMS}
\tsc{BEC}
\tsc{DE}
\usepackage{caption}
\usepackage{lineno}

\usepackage{amsthm}

\newtheorem{assumption}{}
\renewcommand{\theassumption}{A.\arabic{assumption}}

\begin{document}

\let\WriteBookmarks\relax
\def\floatpagepagefraction{1}
\def\textpagefraction{.001}
\shorttitle{Proliferation regulation}
\shortauthors{M.K Gupta et~al.}


\title [mode = title]{Density-dependent growth emerges from Bayesian adaptation of phenotype}

\affiliation[1]{organization={Center for Interdisciplinary Digital Sciences, Department Information Services and High Performance Computing, TUD Dresden University of Technology}, city={Dresden},country={Germany}}

\affiliation[2]{organization={Mathematics Department, Khalifa University, P.O. Box: 127788, Abu Dhabi, UAE}}


\author[1]{Manish Kumar Gupta}[orcid=0009-0003-9649-1894]
\cormark[1]
\ead{manishsvm30@gmail.com}
\credit{Investigation, Conceptualization, Methodology, Algorithm design, Writing – original draft}

\author[2]{Arnab Barua}
[orcid=0000-0002-8210-5978]
\credit{Conceptualization, Writing – review $\And$ editing}

\author[1,2]{Haralampos Hatzikirou}
[orcid=0000-0002-1270-7885]
\ead{haralampos.hatzikirou@ku.ac.ae}
\credit{Investigation, Conceptualization, Supervision, Writing – review $\And$ editing}

\cortext[cor1]{Corresponding author}

\begin{abstract} 
Classical models often describe early tumor expansion as exponential growth, yet experimental and clinical evidence shows that tumor populations can deviate systematically from this behavior, exhibiting density-dependent proliferation, cooperative low-density growth, intermediate growth optima, and finite upper growth bounds before resource limitation or spatial crowding dominate. These observations raise a common question: why should the per-capita growth rate depend on population size?
Here, we propose that sensing mismatch provides a mesoscopic link between environmental change and density-dependent proliferation. We model the cell as a Bayesian adaptive agent whose coarse-grained phenotype evolves on an intrinsic regulatory landscape, while environmental sensing reweights phenotypic states according to how well they account for the extracellular signal statistics generated by the population. In the weak phenotype--signal correlation regime, the stationary phenotype distribution is Gaussian, with its mean displaced from the proliferative optimum by a population-size-dependent baseline information mismatch. This displacement produces a quadratic penalty in the per-capita growth rate.
Coupling the framework to a receptor--ligand decoding model, we show that basal readout error and nonlinear receptor saturation make the mismatch non-monotonic in population size. This single structure gives rise to an intermediate proliferation optimum, an Allee survival threshold, a tissue-specific capacity, and superlinear scaling at low density. A phase diagram in the phenotype--signal coupling and readout-error plane partitions growth into regulated, uncontrolled, and arrested regimes. Thus, density-dependent proliferation need not be imposed phenomenologically, but can emerge from cellular sensing and inference.
\end{abstract}


\maketitle


\section{Introduction}

Understanding how cells regulate proliferation in response to  population size remains a central problem in  developmental biology and cancer research \cite{raff1996size,yankeelov2016multi}. This question is especially important in cancer, where the proliferative behavior of small tumor-cell populations determines whether they collapse, persist, or expand, thereby shaping tumor initiation, recurrence, and metastatic outgrowth \cite{risson2020current,massague2016metastatic,mckenna2018precision}. Classical descriptions of early tumor expansion assume exponential growth, in which each cell proliferates independently and the population growth rate is proportional to the number of cells present. This assumption is mathematically convenient and biologically plausible when cells behave as autonomous proliferative units.

Several recent experimental and clinical observations indicate that tumor growth deviates substantially from this picture. Tumor metabolic activity and proliferation-related imaging signals have been reported to scale superlinearly with tumor volume across multiple human cancers \cite{perez2020universal,bosque2023metabolic,neufeld2017role}, while controlled \textit{in vitro} measurements at low seeding densities show Allee-like behavior, in which the per-capita growth rate increases with population size \cite{johnson2019cancer,panigrahy2012epoxyeicosanoids}. Together with the existence of a finite carrying capacity and intermediate proliferation optima, these phenomena raise a common question: \emph{Why should the per-capita growth rate of a cell population depend on its size, even before classical resource limitation occurs}?

To answer this question, researchers have proposed two broad mechanisms. The first is the evolutionary mechanism: P\'erez-Garc\'ia and colleagues showed that superlinear tumor growth can arise if larger populations accumulate more aggressive subpopulations through mutation or heritable phenotypic change \cite{perez2020universal}.  The second mechanism is signaling-based. Lauffenburger and Cozens showed that inoculum-density effects in mammalian cell culture can emerge from autocrine growth-factor secretion, diffusion and receptor sensing \cite{lauffenburger1989regulation}. Their theory predicts an inoculum density below which growth cannot be sustained. Above this threshold, proliferation increases with population size because denser cultures build up stronger extracellular signals, enhancing the growth stimulus received by each cell.

Although both mechanisms are mechanistically informative, each is specific in a way that limits its generality. The evolutionary route attributes density-dependent growth to changes in population composition through mutation or heritable phenotypic variation. It therefore cannot, on its own, explain density-dependent growth observed in clonal populations over timescales where genetic evolution is not the primary driver \cite{johnson2019cancer}. The autocrine route, by contrast, explains density dependence through a particular signaling implementation, typically involving ligand secretion, receptor binding, and downstream pathways such as MAPK. While powerful, this description must be reformulated when the relevant ligand, receptor system, or intracellular signaling architecture changes. What is missing is a mesoscopic description that captures the common functional structure of these mechanisms without committing to a specific molecular realization.

Although the two mechanisms operate through different biological processes, we propose that they share a common functional structure. At a mesoscopic level of description—above the molecular details of signaling pathways but below macroscopic population dynamics—the key effect of population size is to alter the relationship between the environmental signals available to a cell and the signals it actually senses. We treat this sensing discrepancy as a coarse-grained variable that abstracts away the specific molecular machinery responsible for it.
Under this hypothesis, seemingly distinct mechanisms can be described within a common framework. Population size modifies the information available to individual cells, and cells respond by adjusting their phenotypic state to remain adapted to their environment. This perspective motivates modeling the cell as a Bayesian adaptive agent that updates its internal representation in response to sensory information. We do not suggest that cells literally perform Bayesian inference; rather, Bayesian updating provides a minimal coarse-grained description that captures the information-theoretic structure of sensing-driven adaptation \cite{barua2026bayesian,friston2010free,heins2023collective,auconi2022gradient,mayer2019well,sarkka2023bayesian}.

Within this view, the cell is described by a coarse-grained internal phenotypic state that evolves on an intrinsic regulatory landscape. For a fixed genotype, this landscape encodes the cell's regulatory tendency in the absence of environmental input, with a minimum at the maximally proliferative phenotype. This formulation is consistent with the biological default-state view that proliferation is a constitutive cellular tendency constrained by intracellular regulation rather than induced from zero \cite{waddington2014strategy,huang2005cell,huang2013genetic,soto2016biological}. Autocrine sensing then enters as a Bayesian inference-based update of phenotypic states: phenotypes whose sensed signal statistics better match the true extracellular signal are reinforced, while those that match poorly are depleted.
The mismatch between sensed  and true signal distributions therefore acts as a population-size-dependent bias on the intrinsic landscape, displacing the steady-state phenotype away from the proliferative optimum and reducing the average proliferation rate.

Within this coarse-grained inference framework, a range of population-level growth behaviors arises from a common mechanism rather than being imposed as separate phenomenological assumptions: reduced survival at small population sizes, an intermediate proliferation optimum, a tissue-specific capacity, and superlinear scaling of growth at small population sizes.  
Our work complements rather than displaces evolutionary explanations of non-exponential tumor growth. Mutation may certainly contribute, but density-dependent proliferation can also arise from the way in which a clonal population statistically infers its microenvironment under constraints imposed by internal regulatory architecture

The remainder of this paper is organized as follows. In Sec. 2, we develop the mathematical framework, formalizing phenotypic adaptation as a Bayesian update over an intrinsic regulatory landscape and deriving the stationary phenotype distribution in the weak-correlation regime. Sec. 3 links the resulting representative-cell dynamics to population-level proliferation. It shows how a single effective quantity — the population size–dependent \emph{baseline information mismatch}—  driving  Bayesian inference gives rise to an intermediate proliferation optimum, an Allee threshold, a finite carrying capacity, and superlinear scaling at small population size.
We conclude in Sec. 4 by discussing the biological implications of these results, their relation to existing evolutionary and signaling-based accounts of non-exponential growth, and the limitations of the present framework.

\section{Mathematical framework}
\label{sec:Mathematicalframework}

\begin{table}[htbp]
\centering
\caption{Summary of symbols and their meanings in the model.}
\label{tab:notation_population_model}
\renewcommand{\arraystretch}{1.3}
\begin{tabular}{p{4.5cm} p{10.6cm}}
\hline
\textbf{Symbol / Expression} & \textbf{Description} \\
\hline

$X \in \mathbb{R}$ & Internal phenotypic state of a cell (effective intracellular state). \\

$Y \in \mathbb{R}$ & Microenvironment signal sensed by the cell. \\

$p(X,t)$ & Single-cell phenotype density over accessible  states \(X\) at time \(t\). \\

$\phi(X,t)$ & Population-level phenotype density of cells with internal state $X$ at time $t$. \\

$p(Y\mid X)$ & Sensed distribution of signal $Y$ by a cell in internal phenotypic state $X$, capturing phenotype-dependent noisy sensing. \\

$p(Y,t)=\int p(Y\mid X)\,p(X,t)\,dX$ & Marginal sensed signal distribution induced by the current phenotype density.  \\
$p_{pop}(Y,t)=\int p(Y\mid X)\phi(X,t)\,dX$ & Marginal sensed-signal distribution induced by the current population-level phenotype density.  \\

$q(Y;N)$ & True extracellular signal distribution generated by a population of size \(N\). \\

$f(X)$ & Phenotype-dependent proliferation function. \\

$\bar{f}=\int f(X)\,\phi(X,t)\,dX$ & per-capita growth rate. \\

$\gamma$ & Relaxation rate of intracellular regulatory drift toward the optimal phenotype $X^*$. \\

$D_X$ & Diffusion coefficient in phenotype space, representing stochastic phenotypic fluctuations. \\

$\tau$ & Elementary update timescale for phenotypic adaptation. \\

$\rho$ & Correlation coefficient between phenotype $X$ and sensed signal $Y$. \\

$\Delta$ & Variance-normalized mean mismatch between the true signal distribution $q(Y)$ and the marginal sensed distribution $p(Y)$. \\

$\Delta_0$ & Variance-normalized mean mismatch between the true signal distribution $q(Y)$ and the sensed distribution at $X^*$ $p(Y|X^*)$. \\

\hline
\end{tabular}
\end{table}

\subsection{Model definition and assumptions}

We describe proliferation regulation at the single-cell level using a coarse-grained
internal state $X\in\mathbb{R}$, representing an effective intracellular variable
(e.g.\ MAPK pathway activity) and an external state $Y\in\mathbb{R}$, representing
the local microenvironmental ligand concentration. Although both variables may, in
general, be multidimensional, in this work we restrict attention to the scalar case.
We denote by $p(X,t)$ the probability density over accessible phenotypic states for
a single cell, induced jointly by stochastic intracellular dynamics and
environment-driven inference. The model is built on the following biological
assumptions.

\begin{itemize}

\item  \textbf{Phenotype and intrinsic dynamics.}
For a fixed genotype, cells can occupy distinct intracellular states---for example,
through cell-to-cell variation in MAPK pathway activity or reaction rates---and
thereby exhibit different effective behaviors~\cite{geiler2013details}. We represent
this variability by the coarse-grained internal state $X$. Phenotypic differences
influence both proliferation and environmental sensing.

In the absence of environmental response, the accessible phenotypic states evolve
under an intrinsic regulatory landscape $U_{\mathrm{int}}(X)$ constrained by the
underlying intracellular architecture. We make the following assumption about its form.

\begin{assumption}
 The intrinsic phenotypic landscape is harmonic:
\begin{equation}
U_{\mathrm{int}}(X)
=
\frac{\gamma}{2}(X-X^*)^2,
\end{equation}
where $\gamma>0$ measures the strength of intracellular regulatory confinement
around $X^*$.
\label{ass:harmonic}
\end{assumption}

Here, $X^*$ denotes the \emph{ground state}, toward which the phenotype relaxes in the
absence of environmental response. Single-cell phenotype dynamics in this
landscape are described as stochastic exploration under the deterministic force
generated by $U_{\mathrm{int}}(X)$, together with intracellular fluctuations. The
corresponding density dynamics are
\begin{equation}
\left.
\frac{\partial p(X,t)}{\partial t}
\right|_{\mathrm{int}}
=
\frac{\partial}{\partial X}
\left[
p(X,t)
\frac{\partial U_{\mathrm{int}}(X)}{\partial X}
\right]
+
D_X
\frac{\partial^2 p(X,t)}{\partial X^2},
\end{equation}
where $D_X$ denotes the strength of intracellular phenotypic fluctuations.
Using Assumption~\ref{ass:harmonic}, this becomes
\begin{equation}
\left.
\frac{\partial p(X,t)}{\partial t}
\right|_{\mathrm{int}}
=
\gamma
\frac{\partial}{\partial X}
\Big[
(X-X^*)p(X,t)
\Big]
+
D_X
\frac{\partial^2 p(X,t)}{\partial X^2}.
\label{eq}
\end{equation}
Note that this is a classical Ornstein–Uhlenbeck process.

\item \textbf{Extracellular ligand.}
Cells shape their local microenvironment by secreting extracellular growth signals, namely mitogenic ligands that activate MAPK-related pathways~\cite{gerlee2022autocrine,zhang2014autocrine}. We assume that ligand secretion depends on the local population size $N$, but not explicitly on the phenotypic state $X$. Free ligand is produced at a population-dependent rate, undergoes degradation, and binds reversibly to receptors on the surface of each cell. We describe these processes using the chemical Langevin dynamics of the coupled ligand--receptor system. The complete microscopic formulation is provided in SI Note~1.

To connect the extracellular ligand dynamics to the slower phenotypic dynamics, we assume a separation of timescales:
\begin{assumption}
\label{ass:fastligand}
Ligand dynamics relax much faster than phenotypic adaptation and proliferation \citep{alon2019introduction}.
\end{assumption}
Under this assumption, the extracellular ligand rapidly equilibrates at each fixed population size $N$. Phenotype therefore responds to the stationary statistics of the ligand environment rather than to individual fluctuations along the ligand trajectory. Accordingly, we characterize the extracellular environment by the quasi-steady distribution $q(Y;N)$ of free ligand at fixed $N$.

As derived in SI Note~1, the mean and variance of this distribution are
\begin{equation}
\mu_{Y,q}(N) =\frac{\alpha_Y}{K_d d_Y}
\frac{N}{N+K_N}
\equiv
Y_{\mathrm{max}}
\frac{N}{N+K_N},
\qquad
\sigma_{Y,q}^2(N)=
\frac{\mu_{Y,q}(N)}{K_d}.
\label{eq:ligand_main}
\end{equation}
Here, $Y_{\mathrm{max}}=\alpha_Y/(K_d d_Y)$ denotes the maximal mean ligand level. Thus, the mean extracellular signal increases with population size and saturates at large $N$, while the magnitude of its fluctuations is determined by the corresponding mean ligand abundance.

\item \textbf{Extracellular ligand sensing.}
A cell does not directly observe the extracellular ligand distribution
$q(Y;N)$. Instead, it forms an internal representation of its local
microenvironment through receptor--ligand sensing and downstream signal
processing. We describe this sensing process using a general state-dependent measurement model:
\begin{equation}
    Y
    =
    h_0(X;N)
    +
    h_1(X;N)\eta,
    \label{eq:general_measurement_model}
\end{equation}
where $X$ denotes the phenotypic state of the cell, $h_0(X;N)$ specifies the
mean internally represented signal, and $h_1(X;N)$ determines the magnitude of
the sensing noise. The stochastic variable $\eta \sim \mathcal{N}(0,1)$ is
taken to be standard Gaussian noise.

To obtain an analytically tractable approximation, we expand the measurement
model locally around the intrinsic ground state $X^*$ under the following assumptions.
\begin{assumption}
    The mean response $h_0(X;N)$ is approximated to first order
in $X-X^*$ and the noise amplitude is assumed to vary sufficiently slowly
with phenotype that it may be evaluated at its reference-state value
$h_1(X^*;N)$.
\label{ass:linearizationmeasurment}
\end{assumption}
Under Assumption \ref{ass:linearizationmeasurment}, we can write the measurement model in eq \eqref{eq:general_measurement_model} as:
\begin{equation}
    Y
    \simeq
    h_0(X^*;N)
    +
    \left.
        \frac{\partial h_0}{\partial X}
    \right|_{X^*}
    (X-X^*)
    +
    h_1(X^*;N)\eta.
    \label{eq:linearized_measurement_model}
\end{equation}

The  statistics at the ground  state $X^*$ are not introduced
phenomenologically. Instead, they are determined explicitly from the underlying
receptor--ligand binding dynamics. As derived in SI Note~2, the receptor-level
sensing process yields the Gaussian conditional distribution
\begin{equation}
    p(Y\mid X^*;N)
    =
    \mathcal{N}\!\left(
        \mu_{Y\mid X^*}(N),
        \sigma^2_{Y\mid X^*}(N)
    \right).
    \label{eq:receptorsensing}
\end{equation}
The corresponding mean is
\begin{equation}
    \mu_{Y\mid X^*}(N)
    =
    h_0(X^*;N)
    =
    \mu_{Y,q}(N)
    +
    \delta(N),
    \label{eq:reference_sensing_mean}
\end{equation}
where $\mu_{Y,q}(N)$ denotes the true mean extracellular ligand concentration
and $\delta(N)$ is the receptor-level estimation bias. The reference-state
variance is
\begin{equation}
    \sigma^2_{Y\mid X^*}(N)
    =
    h_1^2(X^*;N),
    \label{eq:reference_sensing_variance}
\end{equation}
which quantifies the uncertainty introduced by receptor-level ligand sensing.
Explicit expressions for $\delta(N)$ and
$\sigma^2_{Y\mid X^*}(N)$ are provided in SI Note~2.

To describe how downstream signaling modifies the internally represented
signal away from the ground state, we express the local sensitivity  in terms of the correlation coefficient $\rho$ and the local
standard deviations $\sigma_X$ and $\sigma_Y$ in phenotype and signal space,
respectively:
\begin{equation}
    \left.
        \frac{\partial h_0}{\partial X}
    \right|_{X^*}
    =
    \rho\,
    \frac{\sigma_Y}{\sigma_X}.
    \label{eq:local_sensing_slope}
\end{equation}
The phenotype-dependent conditional mean therefore follows from Eq \eqref{eq:linearized_measurement_model} as
\begin{equation}
    \mu_{Y\mid X}(N)
    =
    \mu_{Y\mid X^*}(N)
    +
    \rho\,
    \frac{\sigma_Y}{\sigma_X}
    \left(X-X^*\right).
    \label{eq:linearsensing}
\end{equation}

Because the conditional variance is frozen at its reference-state value,
\begin{equation}
    \sigma^2_{Y\mid X}(N)
    \simeq
    \sigma^2_{Y\mid X^*}(N),
    \label{eq:frozen_sensing_variance}
\end{equation}
the resulting phenotype-dependent sensing distribution is
\begin{equation}
    p(Y\mid X;N)
    =
    \mathcal{N}\!\left(
        \mu_{Y\mid X}(N),
        \sigma^2_{Y\mid X^*}(N)
    \right).
    \label{eq:phenotype_dependent_sensing}
\end{equation}

Thus, receptor--ligand binding determines the ground state sensing bias and
uncertainty, whereas downstream signaling
governs the local dependence of the internally represented signal on the
phenotypic state.

\item\textbf{Environmental response as inference.}
    The intrinsic landscape $U_{\mathrm{int}}(X)$ defines the intrinsic phenotypic
    tendency of a cell with fixed genotype. Environmental sensing discussed above modifies this
    tendency through a Bayesian update rule. The true extracellular signal
    distribution is $q(Y;N)$; whereas the marginal sensed distribution under the current
    phenotype density is $p(Y,t)$. A phenotype $X$ is assessed by the likelihood
    ratio $p(Y,t\mid X)/p(Y,t)$, which measures whether phenotype $X$ makes the
    experienced signal $Y$ more or less likely relative to the cell's current
    marginal sensed distribution. Over an elementary adaptation time $\tau$,
    \begin{equation}
        p(X,t+\tau)
        =
        p(X,t)
        \int q(Y;N)
        \frac{p(Y,t\mid X)}{p(Y,t)}
        \,dY.
    \end{equation}
    Phenotypes that better sense signal are reinforced; whereas those that provide a poorer representation are depleted. For the full derivation,
    see SI note 5.

\item \textbf{Single-cell phenotype density dynamics.}
    The full dynamics follow from combining the Bayesian update with the intrinsic
    Fokker--Planck dynamics in the additive manner. Taking the continuous-time limit gives
    \begin{equation}
    \frac{\partial p(X,t)}{\partial t}
    =
    \underbrace{
    \frac{1}{\tau}
    \left[
    \int q(Y;N)\left(\frac{p(Y,t\mid X)}{p(Y,t)}-1\right)dY
    \right]p(X,t)
    }_{\text{Bayesian environmental selection}}
    +
    \underbrace{
    \gamma\frac{\partial}{\partial X}\Big[(X-X^*)\,p(X,t)\Big]
    +
    D_X\frac{\partial^2 p(X,t)}{\partial X^2}
    }_{\text{intrinsic relaxation}}.
    \label{fulldynamics}
    \end{equation}

\item \textbf{Proliferation.}
In the model, the phenotype determines not only sensing but also the cell's
proliferation rate $f(X)$. We assume that a cell's \emph{ground state} $X^*$ is its most proliferative state,
so $f$ peaks at $X^*$.
\begin{assumption}
\label{ass:locality}
$f$ is twice differentiable near $X^*$, so it admits a second-order Taylor
expansion there.
\end{assumption}
\noindent
At the maximum, $f'(X^*)=0$ and $f''(X^*)<0$, so the expansion reduces to
\begin{equation}
    f(X) = f_0 - \alpha\,(X - X^*)^2,
    \label{fitnessform}
\end{equation}
with $f_0 = f(X^*)$ the maximum growth rate and $\alpha = -f''(X^*)/2 > 0$ is the
curvature of the fitness landscape.

\item \textbf{Population phenotype density dynamics.}

The population phenotype density is $\phi(X,t)=n(X,t)/N_T$, with $n(X,t)$ the
number of cells in state $X$ and $N_T(t)=\int n(X,t)\,dX$. Because the cells are
identical and independent, $\phi$ obeys the same single-cell dynamics: the
Bayesian selection and intrinsic relaxation act on it exactly as on the
single-cell density. Replication is the only additional contribution---each cell divides at the rate $f(X)$:
{\small
\begin{equation}
    \frac{\partial \phi(X,t)}{\partial t}
    =
    \underbrace{
    \frac{1}{\tau}
    \left[
    \int q(Y;N)\!\left(\frac{p(Y,t\mid X)}{p_{\mathrm{pop}}(Y,t)}-1\right)dY
    \right]\phi(X,t)
    }_{\text{Bayesian selection}}
    +
    \underbrace{
    \gamma\frac{\partial}{\partial X}\Big[(X-X^*)\,\phi(X,t)\Big]
    +
    D_X\frac{\partial^2 \phi(X,t)}{\partial X^2}
    }_{\text{intrinsic fitness}}
    +
    \underbrace{
    \big[f(X)-\bar{f}\,\big]\,\phi(X,t)
    }_{\text{replicator}},
    \label{popdynamics}
\end{equation}
}
with $\bar{f}(t)=\int \phi(X,t)\,f(X)\,dX$ and
$p_{\mathrm{pop}}=\int p(Y,t|X)\phi(X,t)\,dX$. The Bayesian and intrinsic terms
leave $N_T$ unchanged. Only the replicator term alters population size, with
$\dot{N}_T = N_T\,\bar{f}(t)$. The derivation from single-cell count dynamics
is given in SI note 3).

\end{itemize}

\subsection{Gaussian closure and weak-correlation reduction of the Bayesian inference term}
\label{sec:gaussian_closure}

The phenotype-density dynamics in Eq.~\eqref{popdynamics} contain a nonlocal Bayesian inference term,
\begin{equation}
\int q(Y;N)
\left[
\frac{p(Y\mid X)}
{p_{\mathrm{pop}}(Y,t)}
-1
\right]dY,
\label{eq:bayesian_term_general}
\end{equation}
which quantifies how strongly a phenotype $X$ is reinforced or depleted according to its consistency with the extracellular signal statistics. In its general form, this term depends on the full phenotype distribution through the marginal sensed-signal distribution \(p_{\mathrm{pop}}(Y,t)\). To obtain an analytically tractable reduction, we focus on the biologically relevant regime in which the phenotype distribution remains concentrated around its dominant state and phenotype-dependent changes in the sensed signal are modest.

We first introduce a Gaussian closure for the phenotype density.

\begin{assumption}
\label{ass:laplaceapproximation}
The phenotype density is locally approximated by a Gaussian distribution,
\begin{equation}
p(X,t)
\simeq
\mathcal{N}\left(
X;\mu_X,\sigma_X^2
\right).
\label{eq:gaussian_phenotype_closure}
\end{equation}
\end{assumption}

This closure should be understood as a local approximation around the dominant phenotypic state, rather than as a statement that the complete biological state space is exactly Gaussian. It is natural in the present setting because the intrinsic dynamics confine the phenotype around the ground state \(X^*\), while the environmental response is analyzed as a perturbation of this baseline state.

From Eq.~\eqref{eq:phenotype_dependent_sensing}, the conditional sensed-signal distribution is Gaussian. Its mean depends linearly on \(X\), while its conditional variance is evaluated at the reference phenotype and is therefore independent of (X) to the order considered here. Together with Assumption~\ref{ass:laplaceapproximation}, this implies that the induced joint phenotype--signal distribution is bivariate Gaussian:
\begin{equation}
p(X,Y)=
\mathcal{N}\left(
\begin{bmatrix}
X\\
Y
\end{bmatrix};
\begin{bmatrix}
\mu_X\\
\mu_Y
\end{bmatrix},
\begin{bmatrix}
\sigma_X^2
&
\rho\sigma_X\sigma_Y
\\
\rho\sigma_X\sigma_Y
&
\sigma_Y^2
\end{bmatrix}
\right).
\label{eq:bivariate_gaussian_closure}
\end{equation}
Here, \(\rho\) is the local correlation coefficient between the intracellular phenotype (X) and the sensed signal (Y). It summarizes how strongly a change in phenotype modifies the cell's internal representation of its microenvironment.

For the bivariate Gaussian distribution in Eq.~\eqref{eq:bivariate_gaussian_closure}, the likelihood ratio appearing in Eq.~\eqref{eq:bayesian_term_general} admits the Mehler--Hermite expansion~\cite{lancaster1958structure},
\begin{equation}
\frac{p(Y\mid X)}
{p_{\mathrm{pop}}(Y,t)}=
1
+
\rho z_X z_Y
+
\frac{\rho^2}{2}
\left(z_X^2-1\right)
\left(z_Y^2-1\right)
+
\mathcal{O}(\rho^3),
\label{eq:mehler_hermite_expansion}
\end{equation}
where $
z_X
=
\frac{X-\mu_X}{\sigma_X},
z_Y
=
\frac{Y-\mu_Y}{\sigma_Y}.
$ We next restrict attention to the regime in which downstream phenotype--signal coupling is weak.

\begin{assumption}
\label{ass:smallrhoassumption}
The correlation between phenotype and sensed signal is small ($\rho^2 \ll 1$).
\end{assumption}

Under Assumption~\ref{ass:smallrhoassumption}, Eq.~\eqref{eq:mehler_hermite_expansion} can be truncated at second order:
\begin{equation}
\frac{p(Y\mid X)}
{p_{\mathrm{pop}}(Y,t)}
\simeq
1
+
\rho z_X z_Y
+
\frac{\rho^2}{2}
\left(z_X^2-1\right)
\left(z_Y^2-1\right).
\label{eq:likelihood_ratio_rho2}
\end{equation}
This approximation retains the leading mismatch-dependent response, which biases the phenotype distribution, together with the leading quadratic correction generated by phenotype--signal coupling. Averaging Eq.~\eqref{eq:likelihood_ratio_rho2} over the extracellular signal distribution \(q(Y;N)\) gives
\begin{equation}
\begin{aligned}
&\int q(Y;N)
\left[
\frac{p(Y\mid X)}
{p_{\mathrm{pop}}(Y,t)}
-1
\right]dY
\
 \simeq
\rho z_X
\left\langle z_Y \right\rangle_q
+
\frac{\rho^2}{2}
\left(z_X^2-1\right)
\left(
\left\langle z_Y^2 \right\rangle_q
-1
\right),
\end{aligned}
\label{eq:q_averaged_likelihood_ratio}
\end{equation}
where $\left\langle g(Y)\right\rangle_q:=
\int q(Y;N)g(Y)dY.$ The first two standardized moments of the extracellular signal distribution are
\begin{align}
\left\langle z_Y \right\rangle_q
&=
\frac{\mu_{Y,q}-\mu_Y}{\sigma_Y}
=
-\Delta,
\label{eq:mean_zy_q}\\
\left\langle z_Y^2 \right\rangle_q-1
&=
\frac{\sigma_{Y,q}^2}{\sigma_Y^2}
+
\Delta^2
-1,
\label{eq:second_moment_zy_q}
\end{align}
where $
\Delta
:=
\frac{\mu_Y-\mu_{Y,q}}{\sigma_Y}$
\label{eq:standardized_mismatch} is the variance-normalized mismatch between the internally sensed and actual extracellular signal means. To obtain the leading-order form of the inference term, we consider a rapidly equilibrating extracellular environment and phenotype adaptation close to an adapted state.

\begin{assumption}
\label{ass:narrow_environment}
At fixed population size (N), fluctuations in the extracellular ligand environment are small relative to the uncertainty of the internally sensed signal:
\begin{equation}
\frac{\sigma_{Y,q}^2}{\sigma_Y^2}
\ll 1.
\label{eq:narrow_q_variance}
\end{equation}
\end{assumption}

\begin{assumption}
\label{ass:weak_mismatch}
Phenotype adaptation operates sufficiently close to an adapted state that the variance-normalized difference between the sensed and actual signal means remains small:
\begin{equation}
\lvert \Delta \rvert
=
\left|
\frac{\mu_Y-\mu_{Y,q}}{\sigma_Y}
\right|
\ll 1.
\label{eq:weak_delta}
\end{equation}
\end{assumption}

Under Assumptions~\ref{ass:narrow_environment} and
\ref{ass:weak_mismatch},
\begin{equation}
\left\langle z_Y^2 \right\rangle_q-1
\simeq
-1.
\label{eq:second_moment_reduction}
\end{equation}
Substituting Eqs.~\eqref{eq:mean_zy_q} and
\eqref{eq:second_moment_reduction} into
Eq.~\eqref{eq:q_averaged_likelihood_ratio} yields
\begin{equation}
\int q(Y;N)
\left[
\frac{p(Y\mid X)}
{p_{\mathrm{pop}}(Y,t)}
-1
\right]dY
\simeq
-\rho\Delta z_X-
\frac{\rho^2}{2}
\left(z_X^2-1\right).
\label{eq:bayesian_term_standardized}
\end{equation}

Because \(z_X=(X-\mu_X)/\sigma_X\), the Bayesian inference term reduces to a quadratic polynomial in phenotype space:
\begin{equation}
\int q(Y;N)
\left[
\frac{p(Y\mid X)}
{p_{\mathrm{pop}}(Y,t)}
-1
\right]dY=
b_0^{(q)}
+
b_1^{(q)}X
+
b_2^{(q)}X^2,
\label{eq:bayesian_polynomial}
\end{equation}
with coefficients

\begin{equation}
\boxed{\;
b_0^{(q)}=\frac{\rho\,\Delta\,\mu_X}{\sigma_X}
-\frac{\rho^2}{2}\!\left(\frac{\mu_X^2}{\sigma_X^2}-1\right),
\qquad
b_1^{(q)}=-\frac{\rho\,\Delta}{\sigma_X}+\frac{\rho^2\mu_X}{\sigma_X^2},
\qquad
b_2^{(q)}=-\frac{\rho^2}{2\sigma_X^2},
\;}
\label{eq:b012}
\end{equation}

Substituting \eqref{eq:b012} into \eqref{popdynamics} and using the quadratic
fitness \eqref{fitnessform} gives a polynomial Fokker--Planck equation in which
environmental selection and proliferation reweighting both enter as quadratic
sources,
\begin{equation}
\frac{\partial \phi(X,t)}{\partial t}
=
\underbrace{\frac{1}{\tau}\Big(b_0^{(q)}+b_1^{(q)}X+b_2^{(q)}X^2\Big)\phi(X,t)}_{\text{Bayesian selection}}
+\underbrace{\Big[f_0-\alpha(X-X^*)^2-\bar f\Big]\phi(X,t)}_{\text{replicator}}
+\underbrace{\gamma\,\partial_X\!\big[(X-X^*)\phi(X,t)\big]+D_X\,\partial_X^2 \phi(X,t)}_{\text{intrinsic relaxation}}.
\label{eq:closed_pde}
\end{equation}

\subsubsection*{Moment dynamics and steady-state solution}

The closed PDE \eqref{eq:closed_pde} still requires tracking the full density
$\phi(X,t)$. Since the Gaussian closure of
Assumption~\ref{ass:laplaceapproximation} reduces $\phi$ to its first two
moments, it is sufficient to follow the dynamics of the mean $\mu_X$ and
variance $\sigma_X^2$. Projecting \eqref{eq:closed_pde} onto these two moments,
using $\langle(X-\mu_X)^3\rangle=0$ and $\langle(X-\mu_X)^4\rangle=3\sigma_X^4$,
gives
\begin{align}
\dot\mu_X
&=
\underbrace{-\gamma\,(\mu_X-X^*)}_{\text{intrinsic relaxation}}
\underbrace{-\,2\alpha\,(\mu_X-X^*)\,\sigma_X^2}_{\text{replicator}}
\underbrace{-\,\frac{\rho\,\Delta}{\tau}\,\sigma_X}_{\text{Bayesian selection}},
\label{eq:mean_dyn}\\[6pt]
\dot\sigma_X^2
&=
\underbrace{-2\gamma\,\sigma_X^2+2D_X}_{\text{intrinsic relaxation}}  
\underbrace{-\,2\alpha\,\sigma_X^4}_{\text{replicator}}
\underbrace{-\,\frac{\rho^2}{\tau}\,\sigma_X^2}_{\text{Bayesian selection}}.
\label{eq:var_dyn}
\end{align}

The detailed derivation of the reduced moment dynamics from full density is given in the SI note 4.  Here we focus on the regime in which phenotypic relaxation is fast compared with population growth. We therefore assume a separation of time scales,

\begin{assumption}
\label{ass:prliferationslower}
Phenotypic relaxation is fast compared with population growth. Since \(\gamma\) sets the characteristic relaxation rate of the phenotype dynamics, while \(f_0\) sets the effective rate of population-size change, we assume
\begin{equation}
    f_0 \ll \gamma .
\end{equation}
\end{assumption}

Under this assumption, the phenotype distribution rapidly approaches a quasi-stationary state long before the population size changes appreciably. Consequently, the mean and variance do not need to be followed through their fast transient relaxation. Instead, for each slowly varying population state, we approximate the phenotypic moments by the stationary solutions. Nondimensionalizing time by $\tau$ ($\tilde\gamma=\tau\gamma$,
$\tilde D_X=\tau D_X$, $\tilde\alpha=\tau\alpha$), the variance equation
\eqref{eq:var_dyn} closes on its own at steady state,
\begin{equation}
0=-2\tilde\gamma_{\mathrm{eff}}\,\sigma_X^2+2\tilde D_X-2\tilde\alpha\,\sigma_X^4,
\qquad
\tilde\gamma_{\mathrm{eff}}:=\tilde\gamma+\frac{\rho^2}{2},
\end{equation}
where the second-order Bayesian contribution renormalizes the restoring rate,
$\tilde\gamma\to\tilde\gamma_{\mathrm{eff}}$. The physical root is
\begin{equation}
(\sigma_X^{\mathrm{ss}})^2
=
\frac{-\tilde\gamma_{\mathrm{eff}}
+\sqrt{\tilde\gamma_{\mathrm{eff}}^{\,2}+4\tilde\alpha\tilde D_X}}
{2\tilde\alpha}.
\label{eq:var_ss}
\end{equation}

The mean equation, in contrast, couples to the environment through the mismatch
$\Delta$, which itself depends on $\mu_X$ via the phenotype-dependent sensing
relation~\eqref{eq:phenotype_dependent_sensing}. Closing this self-consistency
gives
\begin{equation}
\Delta(\mu_X,N)
=
\Delta_0(N)+\frac{\rho}{\sigma_X}\,(\mu_X-X^*),
\qquad
\Delta_0(N):=\frac{\mu_{Y|X^*}(N)-\mu_{Y,q}(N)}{\sigma_{Y}},
\label{eq:Delta_closure}
\end{equation}
 where we define the mismatch  $\Delta_0(N)$ at ground state as \emph{baseline mismatch} at  phenotype $X^*$. The full mismatch $\Delta$ adds to this baseline a linear correction
in the displacement $\mu_X-X^*$ induced by the Bayesian feedback. Substituting
\eqref{eq:Delta_closure} into the stationary condition of \eqref{eq:mean_dyn}
and solving for $\mu_X^{\mathrm{ss}}$,
\begin{equation}
\mu_X^{\mathrm{ss}}
=
X^*
-\frac{\rho\,\sigma_X^{\mathrm{ss}}\,\Delta_0(N)}
{\tilde\gamma+\rho^2+2\tilde\alpha\,(\sigma_X^{\mathrm{ss}})^2}.
\label{eq:muX_closed_final}
\end{equation}

\newpage
\section{Results}
\label{sec:results}

The biological question motivating this work is: Why should the
per-capita growth rate of a cell population depend on the size of that population, even in the absence of resource limitation, spatial crowding, or genetic heterogeneity? 
The framework of Sec. 2 was developed to answer this question
without postulating a population size-dependent growth rule. 
It establishes a closed mechanistic link in which the proliferation regulation enters through population size-dependent mismatch rather than through explicit modeling of intracellular regulatory networks.

The four subsections that follow examine what kinds of growth behaviors this framework produces. 
Section~\ref{sec:results-penalty} shows how the effective quantity, the \emph{baseline information mismatch} $\Delta_{0}(N)$, which drives Bayesian update, shapes the  per-capita growth rate $\bar{f}$.
Section~\ref{sec:results-optimum} shows how minimal receptor-level decoding makes $\Delta_{0}(N)$ non-monotonic in population size $N$, thereby predicting an intermediate population size of maximal $\bar{f}$.
Section~\ref{sec:results-allee} identifies the parameter
regime in which the same mechanism generates both an Allee threshold and a tissue-specific capacity. Section~\ref{sec:results-superlinear} shows that  rise of $\bar{f}$ in the low-$N$ regime produces superlinear scaling
$\dot{N}\sim N^{\eta}$ with $\eta>1$, without imposing such scaling as a
phenomenological law.

\subsection{Mismatch suppresses proliferation}
\label{sec:results-penalty}

In the weak-correlation approximation, Sec.~\ref{sec:Mathematicalframework} showed that
the quasi-stationary phenotype density distribution is Gaussian with the mean and variance given in \eqref{eq:muX_closed_final} and \eqref{eq:var_ss},

Crucially, variance of the phenotype distribution is fixed by intrinsic dynamics alone; only the mean carries the dependence on the population size $N$, through the $N$-dependent \emph{baseline mismatch} $\Delta_{0}(N)$, the effective quantity derived from
Bayesian inference in this regime.

The consequence of this shift in mean phenotype for the population-averaged
proliferation rate follows from averaging the proliferation function
$f(X)=f_{0}-\alpha(X-X^{*})^{2}$ over
phenotype distribution. The per-capita growth rate
$\bar{f}(N):=\int f(X)\,\phi_{\text{ss}}(X;N)\,dX$ evaluates to
\begin{equation}
\bar{f}(N)
\;=\; f_{0}
\;-\; \alpha\,(\sigma_X^{\mathrm{ss}})^2
\;-\; \alpha\,
\frac{\rho^2\,(\sigma_X^{\mathrm{ss}})^2}
{\bigl(\tilde\gamma+\rho^2+2\tilde\alpha\,(\sigma_X^{\mathrm{ss}})^2\bigr)^{2}}
\,\Delta_{0}(N)^{2}.
\label{eq:fbar}
\end{equation}
Three physically distinct contributions appear. The bare rate $f_{0}$ sets
the upper bound. The fluctuation correction $\alpha\,(\sigma_X^{\mathrm{ss}})^2$ is
intrinsic: even with perfect sensing, the finite spread $(\sigma_X^{\mathrm{ss}})^2$
of the stationary phenotype distribution costs the population a fixed amount of
fitness, because cells sit, on average, at a distance $\sigma_X^{\mathrm{ss}}$ from
$X^{*}$ on the fitness curvature $\alpha$. The third term is the
inference-induced penalty: it scales quadratically with $\Delta_{0}(N)$ and is
the only channel through which population size enters $\bar{f}$. Its
magnitude combines the displacement of the mean,
$(\mu_{X}^{\text{ss}}-X^{*})^{2}$, together with the same fitness curvature $\alpha$, so
mismatch enters $\bar{f}$ exactly as a coherent shift of the population off
the optimum.

Importantly, the strength of this penalty is governed by the coupling $\rho$ through the
prefactor
$h(\rho)=\rho^2\,(\sigma_X^{\mathrm{ss}})^2/(\tilde\gamma+\rho^2+2\tilde\alpha\,(\sigma_X^{\mathrm{ss}})^2$,
which vanishes at both extremes of $\rho$ and is maximal in between. For $\rho\to0$ --- weak phenotype--signal
coupling --- the cell barely responds to
sensing error, hence the displacement of the mean goes to zero. Interestingly, for large $\rho$, the cell responds sharply to the error, but the same coupling feeds back through $\rho^2$ in the denominator to
resist the resulting displacement, so the realized shift saturates and again
pulls $\mu_{X}^{\text{ss}}$ back toward $X^{*}$. The mismatch-mediated growth
penalty is therefore strongest neither in the weakly nor in the
strongly coupled limit, but at an intermediate \emph{balance point}.

The location of this balance point is set by the
joint $\rho$ dependence of the displacement and the stationary spread
$(\sigma_X^{\mathrm{ss}})^2$, and admits no simple closed form; only in the
weak-selection limit $\tilde\alpha\to0$ does it collapse to the simple condition $\rho_{\mathrm{balance}}^2\approx0.618\,\tilde\gamma$ where the prefactor becomes maximum.

\subsection{Non-monotonic proliferation from density-dependent information mismatch}
\label{sec:results-optimum}
Section~\ref{sec:results-penalty} showed that the per-capita growth rate $\bar{f}$ is penalized quadratically by the \emph{baseline mismatch}
$\Delta_{0}(N)$. Whether this penalty translates into a nontrivial
density dependence of growth depends entirely on how $\Delta_{0}(N)$
itself varies with population size $N$. We therefore establish the
central structural property of $\Delta_{0}(N)$ on which all subsequent results rely.

As discussed previously in the method section, we estimated $\Delta_0(N)$ under receptor-based ligand sensing. We show that $\Delta_0(N)$ exhibits a U-shaped dependence on $N$, with a finite minimum at an intermediate density. The full derivation is given in SI note 2. Briefly, two
opposing sources of decoding error compete as a function of the
$N$-dependent actual mean ligand level $\mu_{Y}^{q}(N) = Y_{\max} N/(N+K_N)$.
At low $N$, ligand is scarce, and the basal readout offset
$\epsilon$ dominates the receptor signal, so the cell cannot reliably
distinguish ligand-driven occupancy from background. At high $N$,
ligand is abundant, and finite-sampling shot noise begins to bound
the decoding accuracy, so the signal-dependent contribution to the
mismatch takes over. The two error sources cross at an intermediate
density. Combining them, the information mismatch is given by
\begin{equation}
    \Delta_0(N)
    \approx
    \sqrt{\frac{n}{R_T\,\mu_{Y}^{q}(N)}}\,\epsilon
    +
    \sqrt{\frac{\mu_{Y}^{q}(N)}{n\,R_T}},
    \qquad
    \mu_{Y}^{q}(N) = Y_{\max}\,\frac{N}{N+K_N},
    \label{deltaNresult}
\end{equation}
with a minimum located at
\begin{equation}
    N^{*} = \frac{n\,\epsilon\,K_N}{Y_{\max} - n\,\epsilon}
\label{eq:Nstar_result}
\end{equation}
where $\epsilon$ is the basal error, $R_T$ is the effective receptor number per cell, and $n$ is the number of independent samples used in receptor-level decoding. Notably, in the noiseless limit,
$\epsilon \to 0$, $N^{*} \to 0$ and the U-shape collapses to a
monotonically increasing curve.

\begin{figure}[pos=tbp!] \centering 
\includegraphics[width=0.5\linewidth]{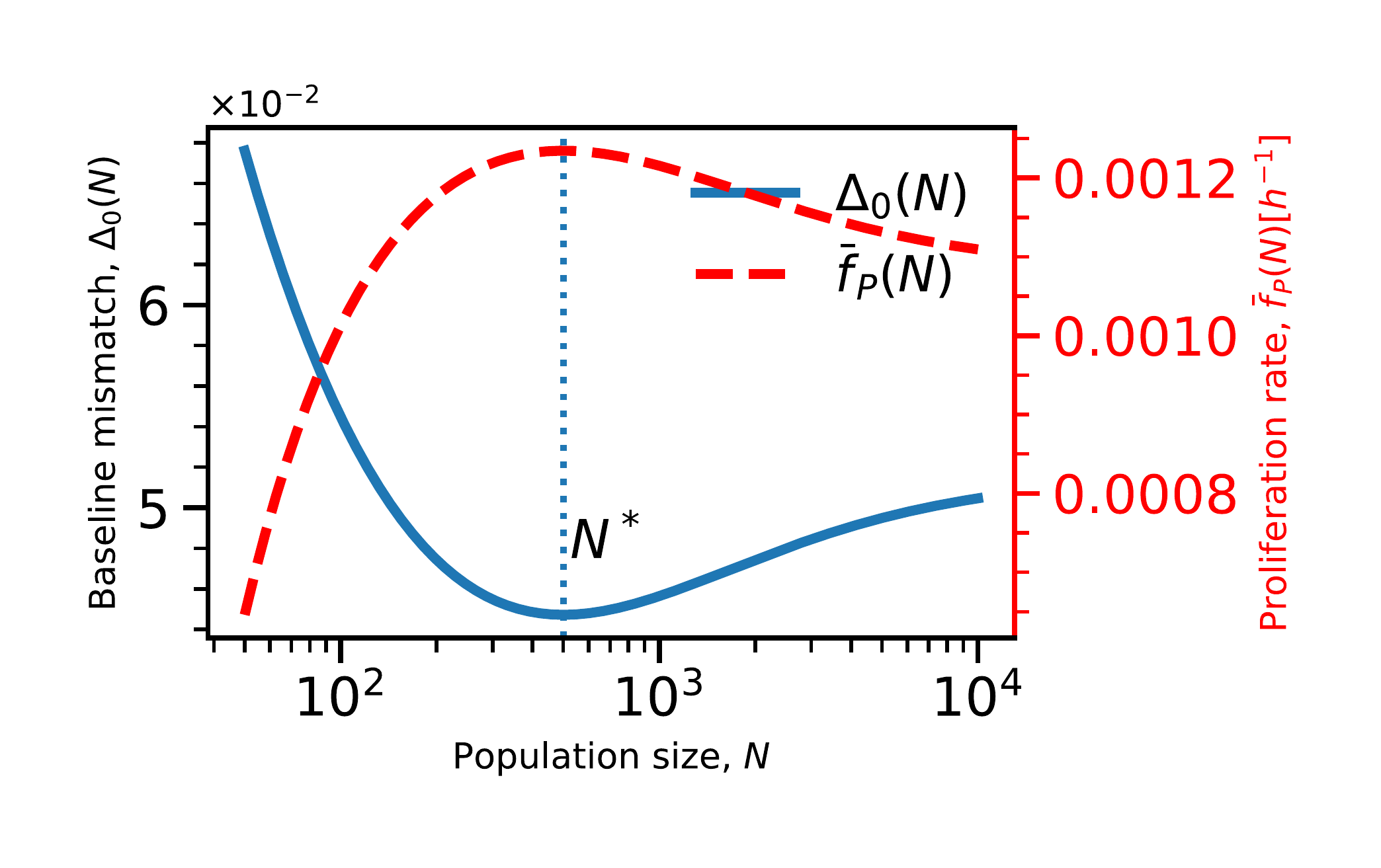} \caption{\textbf{Distinct density-dependent growth regimes.} Mismatch $\Delta_0(N)$ (blue, left axis) and effective proliferation rate (red, right axis) as functions of population size $N$. The dashed vertical line indicates the crossover population size $N^{*}$, separating the positive feedback of density regime $(N<N^{*})$ from the negative density feedback regime $(N>N^{*})$. The parameters used for this figure are given in Table 2 in the supplementary information.} \label{densitydependencedistinctrgime}
\end{figure}

The non-monotonic structure of $\Delta_{0}(N)$ has an immediate
consequence for $\bar{f}(N)$. Since the inference penalty in
Eq.~\eqref{eq:fbar} scales with $\Delta_{0}(N)^{2}$, any local minimum
of the mismatch is automatically a local maximum of the per-capita
proliferation rate. The receptor-level decoding model therefore
predicts a single, finite, intermediate  $N^{*}$ at which the
population reads out its environment most accurately and at which
$\bar{f}$ is maximal (Fig.~\ref{densitydependencedistinctrgime}).
This non-monotonicity reorganizes the $N$-dependence of growth into
two qualitatively distinct regimes separated by $N^{*}$. For $N<N^{*}$,
adding cells improves collective sensing, reduces mismatch, and raises
the per-capita proliferation rate, so feedback from population size is positive
and proliferation is cooperative in form. For $N>N^{*}$, further
increases in $N$ amplify $\Delta_{0}(N)$ and suppress per-capita
proliferation, so feedback from population size is negative, and proliferation is
self-limiting. Crucially, both regimes arise from the same sensing
mechanism: low-population-size facilitation and high-population-size inhibition are
not independent ingredients of the model but two sides of the same
U-shaped decoding error.

\subsection{An Allee threshold and carrying-capacity-like upper bound from information mismatch}\label{sec:results-allee}

The non-monotonicity established in Sec.~\ref{sec:results-optimum} partitions the population space $N$ into cooperative and self-limiting regimes. A more fundamental question is whether the information-mismatch penalty in Eq.~\eqref{eq:fbar} can drive the population-averaged proliferation rate $\bar{f}(N)$ below zero.

In the biologically relevant regime where baseline proliferation outpaces phenotypic variance ($f_{0} - \alpha\,\sigma_X^{2,\mathrm{ss}} > 0$), our model predicts that the average proliferation rate becomes strictly negative whenever the mismatch $\Delta_0(N)$ exceeds a critical value $\Delta_0^{\mathrm{crit}}(\rho)$:
\begin{equation}
\Delta_{0}(N) \;>\; \Delta_0^{\mathrm{crit}}(\rho)
\;:=\;
\frac{\tilde\gamma+\rho^2+2\tilde\alpha\,(\sigma_X^{\mathrm{ss}})^2}{\rho}
\sqrt{\frac{f_{0} - \alpha\,(\sigma_X^{\mathrm{ss}})^2}{\alpha\,(\sigma_X^{\mathrm{ss}})^2}}.
\label{eq:delta_crit}
\end{equation}

\begin{figure}[pos=tbp!] \centering 
\includegraphics[width=0.8\linewidth]{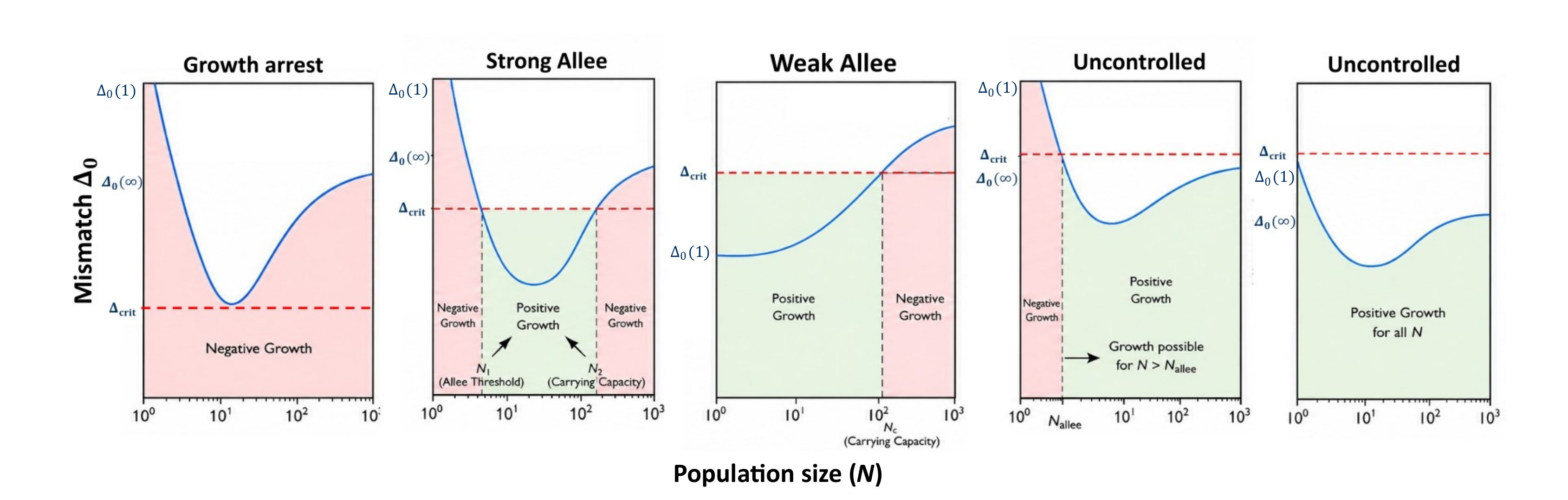} \caption{\textbf{Schematic of growth-regime emergence driven by a non-monotonic baseline information mismatch across population sizes.}}
\label{fig:growth_regimes}
\end{figure}

Crucially, we show five distinct growth regimes can, in principle, emerge. These regimes are determined by the position of the critical mismatch $\Delta_{\mathrm{crit}}(\rho)$ relative to three characteristic values of $\Delta_0(N)$: the low-density value $\Delta_0(1)$, the global minimum $\Delta_0^{\min}$, and the large-$N$ asymptote $\Delta_0^{\infty}$.(Fig.~\ref{fig:growth_regimes}).
\begin{equation}
\Delta_0(1)= \sqrt{\frac{n}{R_T\,\mu_Y^q(1)}}\,\epsilon + \sqrt{\frac{\mu_Y^q(1)}{n R_T}},
\quad
\Delta_0^{\min} = 2\sqrt{\frac{\epsilon}{R_T}},
\qquad
\Delta_0^{\infty} = \sqrt{\frac{n}{R_T\,Y_{\max}}}\,\epsilon + \sqrt{\frac{Y_{\max}}{n R_T}}\,.
\label{eq:delta_min_infty}
\end{equation}

When $\Delta_0^{\mathrm{crit}} < \Delta_0^{\min}$, the critical threshold falls below the entire mismatch curve. Proliferation is negative at every population size, and the population undergoes unconditional extinction regardless of its initial size. We refer to this as the \emph{growth arrest} regime.

When $\Delta_0^{\min} < \Delta_0^{\mathrm{crit}}<\min\{\Delta_0(1),\Delta_0^\infty\}$, the mismatch curve crosses the critical threshold twice, at population sizes
\(N_-<N_+\). Between these two crossings, mismatch is sufficiently low to permit positive proliferation; outside this window, growth is negative. The lower crossing $N_{-}$ functions as an Allee threshold: populations seeded below $N_{-}$ decline toward extinction. The upper crossing $N_{+}$ functions as a tissue-specific capacity: populations that exceed $N_{+}$ experience negative per-capita growth and contract. Growth is therefore confined to a finite density window, and we refer to this as the \emph{strong Allee} regime.

When $\Delta_0(1) < \Delta_0^{\mathrm{crit}} < \Delta_0^{\infty}$, the mismatch at the minimal founder population already lies below the critical level, so there is no lower threshold. However, the curve still crosses $\Delta_0^{\mathrm{crit}}$ once from below at large $N$, producing a single upper bound $N_{+}$. Proliferation is positive for all $N < N_{+}$ and negative above it. We refer to this as the \emph{weak Allee} regime, since low-density facilitation is present --- per-capita growth still increases with $N$ below $N^{*}$ --- but no critical minimum population size is required for survival.

When $\Delta_0^{\mathrm{crit}} > \Delta_0^{\infty}$ and $\Delta_0^{\mathrm{crit}} > \Delta_0(1)$, the threshold lies above the mismatch curve everywhere. Proliferation remains positive at all population sizes, and no upper bound exists. This is the \emph{uncontrolled growth} regime.

Finally, when $\Delta_0^{\mathrm{crit}} > \Delta_0^{\infty}$ but $\Delta_0^{\mathrm{crit}} < \Delta_0(1)$, the threshold exceeds the mismatch curve at large $N$ but falls below it near the minimal founder population. The curve crosses $\Delta_0^{\mathrm{crit}}$ once at low density, producing a single Allee threshold $N_{-}$ with no upper bound. Populations below $N_{-}$ collapse; those above it grow without limit. This constitutes a second variant of the \emph{uncontrolled growth} regime, distinguished from the previous case by the presence of a finite founder-size requirement.\\

The classification above establishes the distinct growth behaviors that can, in principle, arise from the non-monotonic information mismatch. We next ask which combinations of model parameters realize each regime and how variations in the underlying biophysical parameters drive transitions between them. To address this question, we map the resulting growth regimes across the $(\rho, \epsilon)$ parameter space and examine how the critical population sizes $N_{-}$ and $N_{+}$ emerge, shift, or disappear.

\begin{figure}[pos=tbp!] \centering 
\includegraphics[width=0.75\linewidth]{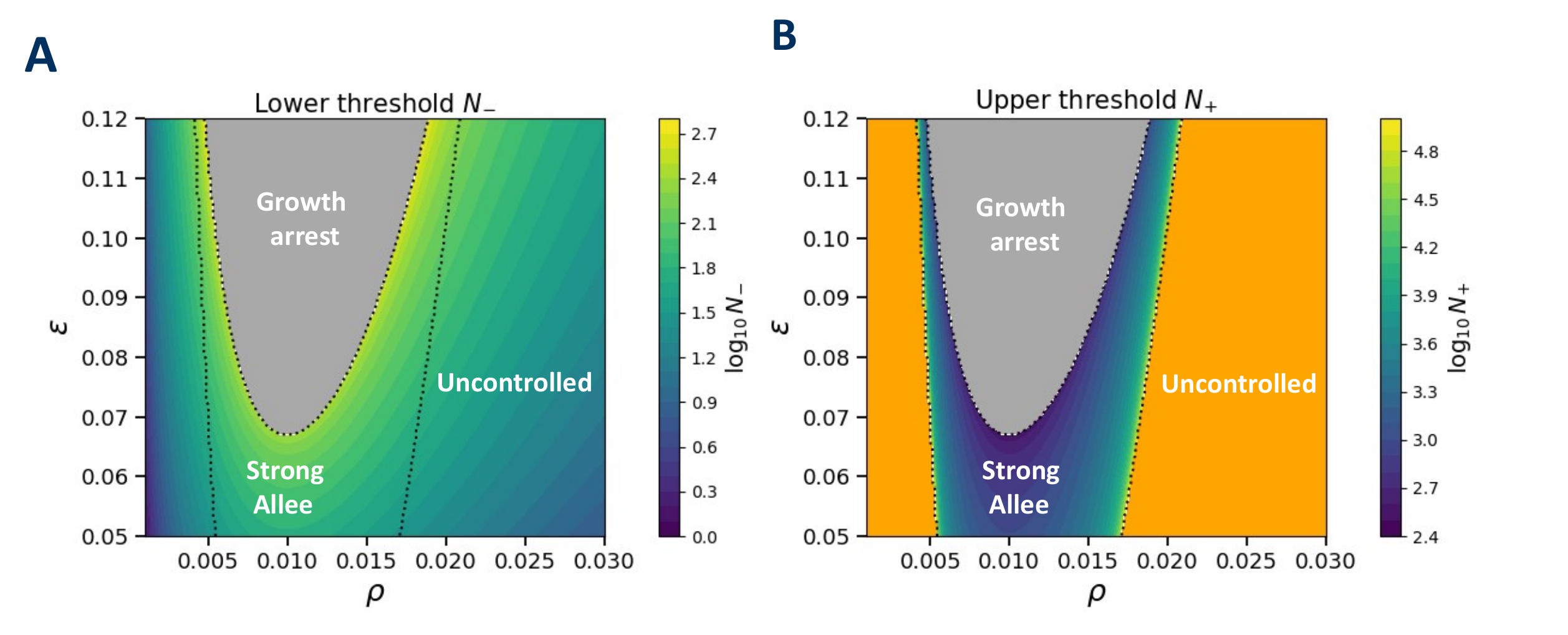} \caption{\textbf{Phase diagram of the critical population sizes.}
\textbf{(A)}~Heatmap of the lower threshold $N_{-}$ (Allee threshold) on a
$\log_{10}$ color scale as a function of mismatch sensitivity $\rho$ and
basal readout error $\epsilon$. The colored region corresponds to parameter
combinations where a finite Allee threshold exists. The dark gray region at large
$\epsilon$ marks the \emph{growth arrest} regime, where
$\Delta_{\mathrm{crit}} < \Delta_0^{\min}$ and proliferation is negative at
every population size.
\textbf{(B)}~Heatmap of the upper threshold $N_{+}$ (carrying capacity) on a
$\log_{10}$ color scale over the same $(\rho, \epsilon)$ plane. The colored
region identifies parameter combinations that admit a finite carrying
capacity.  The orange region marks
the lower-threshold growth regime, where only an Allee threshold exists
but no upper bound constrains the population. The dotted
black boundary separates these qualitatively distinct regimes. The parameter values used for the analysis are provided
in Table~2 of the supplementary information.} \label{heatmap}
\end{figure}

The phase diagram in Fig.~\ref{heatmap} resolves both parameter
dependencies simultaneously. Along the $\rho$-axis, at fixed $\epsilon$,
regulated growth is confined to an intermediate band of
phenotype--signal coupling: outside this band, the carrying capacity
$N_{+}$ diverges and growth becomes unbounded at high density. Within
the band, the boundaries of the growth-permitting window depend
sensitively on $\rho$, and the window itself is most tightly confined
at an intermediate coupling, where $N_{-}$ attains its maximum and $N_{+}$ its
minimum. Moving away from this point in either direction broadens this
window — $N_{-}$ decreases toward zero and $N_{+}$ grows — so the
\textit{balance point} between environmental responsiveness and intracellular
restoration identified in Sec.~\ref{sec:results-penalty} as the
location of maximal mismatch sensitivity is also the location of
maximal population-level growth control.

Along the $\epsilon$-axis, at fixed $\rho$, the offset $\epsilon$ acts
as the regulatory amplitude: increasing $\epsilon$ raises the entire
mismatch curve, simultaneously increasing $N_{-}$ and decreasing
$N_{+}$. The growth-permitting window, therefore, narrows continuously
as the basal readout error grows, and beyond a critical value of
$\epsilon$ the two thresholds collide and disappear, recovering the
growth-arrest regime in which $\Delta_{0}(N) > \Delta_0^{\mathrm{crit}}$
for every $N$. Together, the two axes thus identify $\rho$ as the
parameter that selects the \emph{regime} of growth control --- regulated,
uncontrolled, or arrested --- and $\epsilon$ as the parameter that
sets, within the regulated regime, the \emph{width} of the
density window in which the population can sustain positive growth.

\subsection{Superlinear scaling at low density from density-dependent information mismatch}\label{sec:results-superlinear}


We finally asked whether the Bayesian inference-based mechanism that produces population-size dependent growth can also reproduce the superlinear growth scaling reported in tumor populations. From the population equation
\begin{equation}
    \dot{N}=N\bar{f}(N),
\end{equation}
exponential growth would correspond to a constant per-capita growth rate, such that \(\dot{N}\propto N\) and the slope on a log--log plot would be one. In our model, however, $\bar{f}(N)$ rises with $N$ in the low-$N$ regime below the optimal sensing size $N^{*}$ because $\Delta_{0}(N)$ falls and the quadratic penalty in Eq.~\eqref{eq:fbar} weakens. Thus, the model predicts that $\dot{N}$ grows faster than linearly with \(N\).

\begin{figure}[pos=tb!] \centering 
\includegraphics[width=0.9\linewidth]{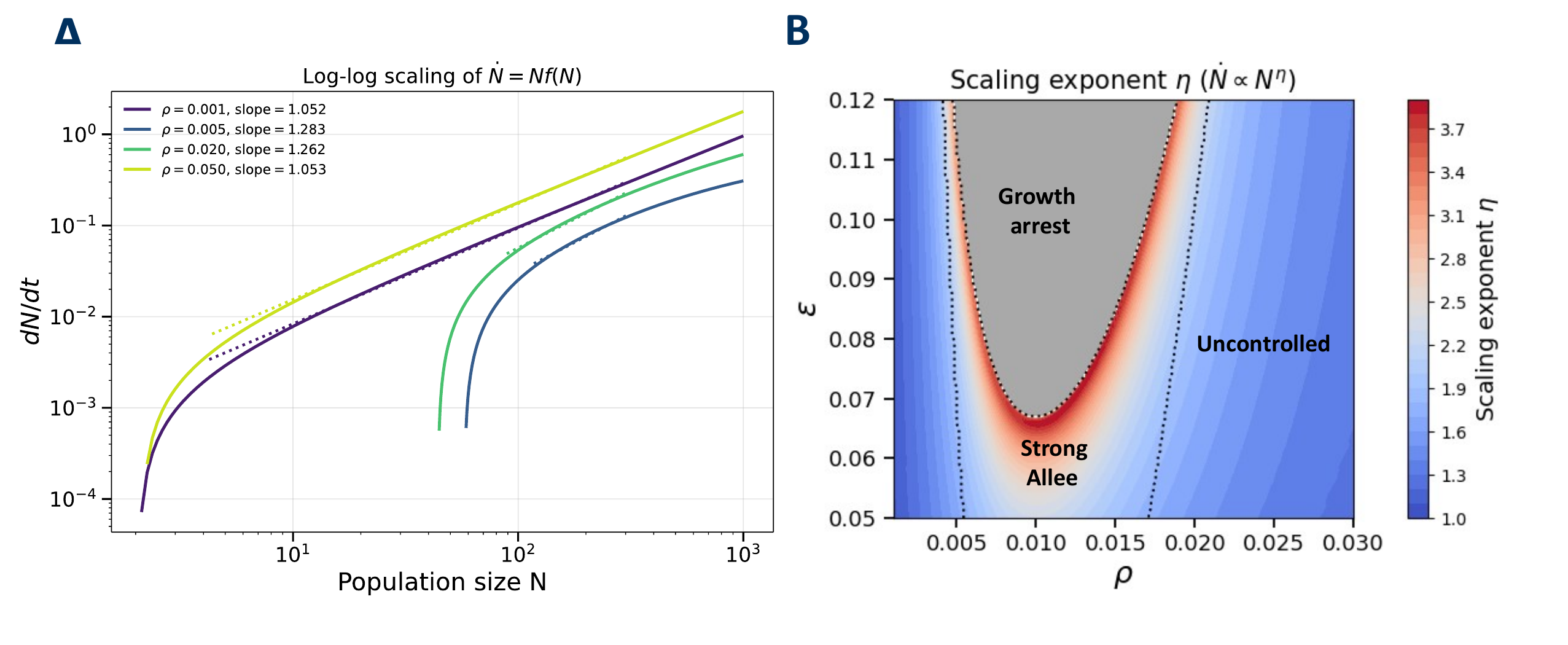} \caption{\textbf{Log--log scaling of the population growth rate under mismatch-driven proliferation regulation.}
\textbf{A,} Population growth rate \(\dot{N}=dN/dt=N\bar{f}_{P}(N)\) plotted against population size \(N\) on log--log axes for different values of correlation $\rho$. Solid curves show the theoretical prediction from the mismatch-driven growth law, and dotted lines show power-law fits over the low-density growth-permitting regime.   Fits are performed over the low-density growth-permitting interval, using
\(2N_-\) as the lower bound when an Allee threshold exists, and the smallest
positive simulated population size otherwise.
The fitted slopes give the effective scaling exponent \(\eta\) in \(\dot{N}\sim N^{\eta}\). Slopes greater than one indicate superlinear growth, showing that the per-capita proliferation rate increases with population size in the low-density regime.
\textbf{B,} Heatmap of the fitted local scaling exponent \(\eta\) across the \((\rho,\epsilon)\) parameter plane. Superlinear growth occurs throughout the displayed regime, but the exponent is largest in the regulated-growth region, where mismatch-mediated feedback produces both an Allee threshold and a carrying capacity. The dashed boundary separates regulated growth from uncontrolled growth. The parameters used for this figure are given in Table 2 of the supplementary information.}\label{scalinglaw}
\end{figure}

Across the explored range of the mismatch-sensitivity parameter $\rho$, the theoretical
$\dot{N}$ curves lie systematically above the unit-slope
exponential reference over the low-density growth-permitting window,
and a power-law fit
\begin{equation}
\dot{N} \;\sim\; N^{\eta},
\qquad
\eta > 1,
\label{eq:powerlaw}
\end{equation}
captures the local scaling (Fig.\ref{scalinglaw} A). The exponent $\eta$ is therefore not an
input to the model but an emergent readout of how sharply
$\bar{f}$ rises with $N$ at low density.

The non-monotonic dependence of $\eta$ on $\rho$ arises from the nonlinear
$\rho$-dependence of the prefactor multiplying the mismatch penalty in
Eq.~\eqref{eq:fbar}. For small $\rho$, the phenotype is only weakly coupled
to the microenvironment, so environmental changes produce only a small
variation in the population-averaged fitness $\bar{f}$, keeping $\eta$ close
to unity. For large $\rho$, the phenotype is highly sensitive to the
microenvironment, so a given environmental change can be accommodated by
only a small phenotypic displacement. The resulting mismatch penalty is
therefore again small, and $\eta$ approaches unity. Consequently, the
strongest reduction in $\eta$ occurs at intermediate $\rho$. The phase diagram in Fig.~\ref{scalinglaw}B
confirms this across the $(\rho, \epsilon)$ plane: $\eta > 1$ holds
throughout the displayed regime, with the strongest superlinearity
concentrated inside the regulated-growth region.

The framework therefore reproduces, from a single
information-theoretic mechanism, the empirical low-density scaling
that motivated the introduction. Low-density populations sense their
environment poorly, the phenotype is displaced from $X^{*}$, and
proliferation is correspondingly suppressed; as the population grows,
sensing improves, mismatch falls, and the per-capita rate rises. The
resulting $\dot{N} \sim N^{\eta}$ scaling is not a phenomenological
growth law fitted to data but an emergent consequence of how a clonal
population infers its own density --- complementing, rather than
replacing, evolutionary explanations that attribute the same scaling
to subclonal heterogeneity.

\section{Discussion}
\label{sec:discussion}
Understanding why tumor populations deviate from autonomous exponential growth  is biologically important, as it can strongly influence tumor initiation, persistence, and recurrence. A major challenge is the limited understanding of how sensory signals received at the tumor-cell surface are interpreted by intracellular regulatory networks to regulate proliferation.
Here, we bypass the need for detailed mechanistic knowledge by modeling the cell as a Bayesian adaptive agent whose phenotype evolves on an intrinsic regulatory landscape. Autocrine growth-factor sensing provides an external information-driven bias that displaces the effective landscape minimum away from the proliferation optimum. This bias is generated by the mismatch between the true extracellular signal distribution, $q(Y)$, and the signal distribution sensed by the cell (via receptor mediated sensing and downstream signaling), $p(Y)$, where $p(Y)$ is induced by the current phenotypic distribution through $p(Y)=\int p(Y|X)p(X)\,dX$.

In this view, growth is reduced when sensing mismatch displaces the steady-state phenotype away from the proliferation-optimal state, and the magnitude of this displacement is determined not only by the mismatch itself but also by the phenotype--signal correlation, which acts as a multiplicative prefactor. Importantly, the prefactor does not regulate mismatch sensitivity monotonically. Instead, the model predicts that the susceptibility of the phenotype to mismatch-driven deviation from the optimum is maximal at an intermediate coupling strength, whereas both weak and excessively strong coupling reduce this effect. Biologically, this means that proliferation is most tightly controlled when environmental information is coupled strongly enough to influence phenotype, but not so strongly that the feedback effectively saturates. This non-monotonic dependence on  phenotype-signal correlation has a direct implication for understanding heterogeneity in density sensitivity across cell populations. Cell lines or tumor types that differ in their effective phenotype-signal coupling — through differences in receptor expression, downstream signaling gain, or chromatin accessibility — are predicted to differ systematically in how strongly their proliferation responds to changes in local density. Populations near the intermediate coupling regime should show the strongest density dependence, while those at either extreme should approach density-independent growth. This provides a mechanistic interpretation of the empirically observed variability in density-dependent growth behavior across cancer cell lines, which has previously been treated as a cell-line-specific phenomenological parameter rather than a consequence of a shared sensing mechanism.

Population-size-dependent mismatch emerging through receptor-mediated decoding reproduces observed growth behaviors: low-density growth promotion, high-density inhibition, an intermediate density of maximal proliferation, Allee-like survival thresholds, and tissue-specific capacity. It also provides an interpretation of the superlinear scaling of population growth with population size in the low-density regime as a consequence of mismatch reduction with increasing population size, thereby linking apparent allometric growth laws to the underlying phenotype-signal correlation parameter. Beyond reproducing these individual phenomena, the framework predicts a phase structure in parameter space that partitions cell populations into three qualitatively distinct growth regimes: \textit{regulated growth}, in which both an Allee threshold and a tissue specific capacity exist and the population can persist only within a finite density window; \textit{uncontrolled growth}, in which an Allee threshold exists but there is no upper bound on population growth; and \textit{growth arrest}, in which mismatch exceeds the critical level and population decays. This trichotomy has a direct biological interpretation. Regulated growth corresponds to normal density-dependent control in which collective sensing keeps the population within viable bounds \cite{martin1997wound,penzo2015organ,gokhale2015size}. The transition to uncontrolled growth corresponds to a loss of the upper regulatory threshold while the lower threshold persists — which is precisely the phenomenology of early tumor escape: a population that was previously bounded becomes unbounded not because it proliferates faster at all densities, but because the mismatch mechanism that previously imposed an upper limit has been disrupted. In our framework, this transition occurs when the phenotype-signal coupling $\nu$ moves outside the intermediate regime of balanced sensing and regulatory control and the  phenotype decouples from environmental readout. This resembles one of the most pronounced hallmarks of cancer, which is independence from environmental growth signal \cite{hanahan2000hallmarks,hanahan2011hallmarks,hoxhaj2020pi3k}. 

The Bayesian adaptation equation derived here is not specific to autocrine ligand sensing. Because the Fokker-Planck-selection structure depends only on the mismatch between the cell's predicted signal distribution and the actual environmental signal distribution, the same framework applies in principle to any sensing modality for which the population-size-dependent signal statistics $q(Y;N)$ can be specified. Mechanosensing provides a natural and biologically important example: if $Y$ represents a mechanical contact readout, such as membrane tension or mechanical pressure. Such  a mechanism can generate a density-dependent mismatch that displaces the phenotype away from the proliferative optimum through the same Bayesian adaptation mechanism. This is contact inhibition, arising from cell sensing rather than the classical geometric exclusion. We therefore expect that a mechanosensing specification of the framework would recover effective contact-inhibition growth laws of the kind derived by Kimmel et al. \cite{kimmel2026universal}, from microscopic spatial exclusion assumptions. This connection suggests that the present framework may serve as a unifying probabilistic structure for density-dependent growth regulation across signal modalities and density regimes, with autocrine ligand sensing governing the low-density regime studied here and mechanosensing governing the high-density contact-inhibition regime. A full derivation of the mechanosensing case, including the precise mapping between the statistical structure of $q(Y;N)$ and the growth laws that emerge, is deferred to future work.

While this framework unifies several density-dependent phenomena, it relies on simplifying assumptions that define clear avenues for future work. First, by treating the external signal as a unidimensional variable (e.g., a single growth factor), we omit the multidimensional nature of the microenvironment. Extending the inference problem to a vector-valued setting would reveal how cross-correlations among nutrients, oxygen, and paracrine factors renormalize proliferation dynamics. 
Second, the current well-mixed mean-field approximation neglects spatial heterogeneity in ligand concentration; relaxing this requires coupling the Bayesian dynamics to spatial transport models. Third, we assume a separation of timescales where phenotypic adaptation equilibrates much faster than population growth. This may break down near bifurcation points, such as the Allee threshold, where transient dynamics dominate. Finally, our quantitative predictions depend heavily on the nonzero basal readout offset ($\epsilon$), and the sensitivity of the phase boundaries to this parameter warrants systematic exploration.

A direct validation of our theory would require parallel measurements of the extracellular ligand concentration and the corresponding receptor binding readouts across cell densities. Comparing these two readouts across population density would allow one to reconstruct the effective sensing mismatch and test the central prediction that this mismatch varies non-monotonically with density. A second level of validation would be verifying if proliferation decreases with a mismatch. If the model is correct, conditions that increase mismatch should shift the population away from the growth-optimal phenotype and reduce the proliferation rate. In this way, the theory predicts a concrete chain of observables: density alters sensing statistics, altered sensing statistics generate mismatch, mismatch displaces phenotype, and phenotype displacement penalizes growth. These links provide a clear experimental route for testing whether the emergence of density-dependent proliferation can indeed be explained through inference error rather than imposed phenomenological growth rules. A third level of validation would explore the phase structure of the model. The regulated-to-uncontrolled transition should be experimentally accessible by manipulating the effective phenotype-signal coupling through titration of receptor expression or downstream signaling components. If this phase boundary can be identified experimentally and shown to correspond to a change in coupling strength rather than a change in intrinsic proliferative capacity, it would provide direct support for the reframing of dysregulation as a sensing failure proposed above.

In conclusion, the present framework identifies sensing mismatch as the common mechanistic principle underlying density-dependent proliferation regulation, with the Bayesian adaptation equation providing a unifying mathematical structure that is in principle applicable across different signal modalities. In the autocrine-ligand application presented here, four distinct low-density phenomena — Allee thresholds, intermediate proliferation optima, tissue-specific capacities, and superlinear growth scaling — emerge as consequences of a single sensing mechanism. The expected connection to contact-inhibition-based growth laws through a mechanosensing extension further suggests that what have previously appeared to be distinct physical mechanisms operating in distinct density regimes may be special cases of a single probabilistic principle of inference-driven phenotypic adaptation. Within this view, density-dependent growth need not be imposed phenomenologically, but can arise naturally from the structure of cellular sensing and mismatch. This provides a principled and potentially general route for linking single-cell inference to population-level growth behavior, and opens a concrete research program for understanding how cells integrate multiple sensing modalities to regulate proliferation in development and disease.

\printcredits

\section*{Declaration of competing interest}
The authors declare no competing interests.

\section*{Acknowledgments and funding}
HH and  MKG would like to thank Volkswagenstiftung for its support of the "Life?"
program (96732). Finally, HH acknowledges the support of the RIG-2025-001 grant from Khalifa University
and the UAE-NIH Collaborative Research grant AJF-NIH-25-KU.

\section*{Data availability}
This paper does not report either data generation or analysis.

\section*{Code availability}
All computational analyses and modeling code supporting this study are available upon request.

\bibliographystyle{unsrtnat}
\bibliography{bibliography}

@article{raff1996size,
  title={Size control: the regulation of cell numbers in animal development},
  author={Raff, Martin C},
  journal={Cell},
  volume={86},
  number={2},
  pages={173--175},
  year={1996},
  publisher={Elsevier}
}

@article{mckenna2018precision,
  title={Precision medicine with imprecise therapy: computational modeling for chemotherapy in breast cancer},
  author={McKenna, Matthew T and Weis, Jared A and Brock, Amy and Quaranta, Vito and Yankeelov, Thomas E},
  journal={Translational oncology},
  volume={11},
  number={3},
  pages={732--742},
  year={2018},
  publisher={Elsevier}
}

@article{yankeelov2016multi,
  title={Multi-scale modeling in clinical oncology: opportunities and barriers to success},
  author={Yankeelov, Thomas E and An, Gary and Saut, Oliver and Luebeck, E Georg and Popel, Aleksander S and Ribba, Benjamin and Vicini, Paolo and Zhou, Xiaobo and Weis, Jared A and Ye, Kaiming and others},
  journal={Annals of biomedical engineering},
  volume={44},
  number={9},
  pages={2626--2641},
  year={2016},
  publisher={Springer}
}

@article{risson2020current,
  author  = {Risson, Emilie and Nobre, Ana R. and Maguer-Satta, V{\'e}ronique and Aguirre-Ghiso, Julio A.},
  title   = {The Current Paradigm and Challenges Ahead for the Dormancy of Disseminated Tumor Cells},
  journal = {Nature Cancer},
  volume  = {1},
  pages   = {672--680},
  year    = {2020},
  doi     = {10.1038/s43018-020-0088-5}
}

@article{massague2016metastatic,
  author  = {Massagu{\'e}, Joan and Obenauf, Anna C.},
  title   = {Metastatic Colonization by Circulating Tumour Cells},
  journal = {Nature},
  volume  = {529},
  number  = {7586},
  pages   = {298--306},
  year    = {2016},
  doi     = {10.1038/nature17038}
}

@article{bosque2023metabolic,
  title={Metabolic activity grows in human cancers pushed by phenotypic variability},
  author={Bosque, Jesus J and Calvo, Gabriel F and Molina-Garc{\'\i}a, David and P{\'e}rez-Beteta, Juli{\'a}n and Vicente, Ana M Garc{\'\i}a and P{\'e}rez-Garc{\'\i}a, V{\'\i}ctor M},
  journal={Iscience},
  volume={26},
  number={3},
  year={2023},
  publisher={Elsevier}
}

@article{panigrahy2012epoxyeicosanoids,
  title={Epoxyeicosanoids stimulate multiorgan metastasis and tumor dormancy escape in mice},
  author={Panigrahy, Dipak and Edin, Matthew L and Lee, Craig R and Huang, Sui and Bielenberg, Diane R and Butterfield, Catherine E and Barn{\'e}s, Carmen M and Mammoto, Akiko and Mammoto, Tadanori and Luria, Ayala and others},
  journal={The Journal of clinical investigation},
  volume={122},
  number={1},
  pages={178--191},
  year={2012},
  publisher={American Society for Clinical Investigation}
}

@article{neufeld2017role,
  title={The role of Allee effect in modelling post resection recurrence of glioblastoma},
  author={Neufeld, Zoltan and von Witt, William and Lakatos, Dora and Wang, Jiaming and Hegedus, Balazs and Czirok, Andras},
  journal={PLoS computational biology},
  volume={13},
  number={11},
  pages={e1005818},
  year={2017},
  publisher={Public Library of Science San Francisco, CA USA}
}

@book{waddington2014strategy,
  title={The strategy of the genes},
  author={Waddington, Conrad Hal},
  year={2014},
  publisher={Routledge}
}

@article{huang2005cell,
  title={Cell fates as high-dimensional attractor states of a complex gene regulatory network},
  author={Huang, Sui and Eichler, Gabriel and Bar-Yam, Yaneer and Ingber, Donald E},
  journal={Physical review letters},
  volume={94},
  number={12},
  pages={128701},
  year={2005},
  publisher={APS}
}

@article{huang2013genetic,
  title={Genetic and non-genetic instability in tumor progression: link between the fitness landscape and the epigenetic landscape of cancer cells},
  author={Huang, Sui},
  journal={Cancer and Metastasis Reviews},
  volume={32},
  number={3},
  pages={423--448},
  year={2013},
  publisher={Springer}
}

@article{friston2010free,
  title={The free-energy principle: a unified brain theory?},
  author={Friston, Karl},
  journal={Nature reviews neuroscience},
  volume={11},
  number={2},
  pages={127--138},
  year={2010},
  publisher={Nature publishing group}
}

@article{heins2023collective,
  title={Collective behavior from surprise minimization. arXiv},
  author={Heins, C and Millidge, B and Da Costa, L and Mann, R and Friston, K and Couzin, ID},
  journal={arXiv preprint arXiv:2307.14804},
  year={2023}
}

@article{barua2026bayesian,
  title={Bayesian Decision-Making Shapes Phenotypic Landscapes from Differentiation to Cancer},
  author={Barua, Arnab and Hatzikirou, Haralampos},
  journal={Entropy},
  volume={28},
  number={3},
  pages={312},
  year={2026},
  publisher={MDPI}
}

@book{sarkka2023bayesian,
  title={Bayesian filtering and smoothing},
  author={S{\"a}rkk{\"a}, Simo and Svensson, Lennart},
  volume={17},
  year={2023},
  publisher={Cambridge university press}
}

@article{mayer2019well,
  title={How a well-adapting immune system remembers},
  author={Mayer, Andreas and Balasubramanian, Vijay and Walczak, Aleksandra M and Mora, Thierry},
  journal={Proceedings of the National Academy of Sciences},
  volume={116},
  number={18},
  pages={8815--8823},
  year={2019},
  publisher={National Academy of Sciences}
}

@article{auconi2022gradient,
  title={Gradient sensing in Bayesian chemotaxis},
  author={Auconi, Andrea and Novak, Maja and Friedrich, Benjamin M},
  journal={Europhysics letters},
  volume={138},
  number={1},
  pages={12001},
  year={2022},
  publisher={EDP Sciences, IOP Publishing and Societ{\`a} Italiana di Fisica}
}

@article{perez2020universal,
  title={Universal scaling laws rule explosive growth in human cancers},
  author={P{\'e}rez-Garc{\'\i}a, V{\'\i}ctor M and Calvo, Gabriel F and Bosque, Jes{\'u}s J and Le{\'o}n-Triana, Odelaisy and Jim{\'e}nez, Juan and P{\'e}rez-Beteta, Juli{\'a}n and Belmonte-Beitia, Juan and Valiente, Manuel and Zhu, Luc{\'\i}a and Garc{\'\i}a-G{\'o}mez, Pedro and others},
  journal={Nature physics},
  volume={16},
  number={12},
  pages={1232--1237},
  year={2020},
  publisher={Nature Publishing Group UK London}
}

@article{johnson2019cancer,
  title={Cancer cell population growth kinetics at low densities deviate from the exponential growth model and suggest an Allee effect},
  author={Johnson, Kaitlyn E and Howard, Grant and Mo, William and Strasser, Michael K and Lima, Ernesto ABF and Huang, Sui and Brock, Amy},
  journal={PLoS biology},
  volume={17},
  number={8},
  pages={e3000399},
  year={2019},
  publisher={Public Library of Science San Francisco, CA USA}
}

@article{lauffenburger1989regulation,
  title={Regulation of mammalian cell growth by autocrine growth factors: analysis of consequences for inoculum cell density effects},
  author={Lauffenburger, D and Cozens, C},
  journal={Biotechnology and bioengineering},
  volume={33},
  number={11},
  pages={1365--1378},
  year={1989},
  publisher={Wiley Online Library}

}

@article{soto2016biological,
  title={The biological default state of cell proliferation with variation and motility, a fundamental principle for a theory of organisms},
  author={Soto, Ana M and Longo, Giuseppe and Mont{\'e}vil, Ma{\"e}l and Sonnenschein, Carlos},
  journal={Progress in Biophysics and Molecular Biology},
  volume={122},
  number={1},
  pages={16--23},
  year={2016},
  publisher={Elsevier}
}

@article{lancaster1958structure,
  title={The structure of bivariate distributions},
  author={Lancaster, Henry Oliver},
  journal={The Annals of Mathematical Statistics},
  volume={29},
  number={3},
  pages={719--736},
  year={1958},
  publisher={JSTOR}
}

@book{alon2019introduction,
  title={An introduction to systems biology: design principles of biological circuits},
  author={Alon, Uri},
  year={2019},
  publisher={Chapman and Hall/CRC}
}

@article{geiler2013details,
  title={The details in the distributions: why and how to study phenotypic variability},
  author={Geiler-Samerotte, KA and Bauer, CR and Li, S and Ziv, N and Gresham, David and Siegal, ML},
  journal={Current opinion in biotechnology},
  volume={24},
  number={4},
  pages={752--759},
  year={2013},
  publisher={Elsevier}
}

@article{zhang2014autocrine,
  title={Autocrine VEGF signaling promotes proliferation of neoplastic Barrett's epithelial cells through a PLC-dependent pathway},
  author={Zhang, Qiuyang and Yu, Chunhua and Peng, Sui and Xu, Hao and Wright, Ellen and Zhang, Xi and Huo, Xiaofang and Cheng, Edaire and Pham, Thai H and Asanuma, Kiyotaka and others},
  journal={Gastroenterology},
  volume={146},
  number={2},
  pages={461--472},
  year={2014},
  publisher={Elsevier}
}

@article{gerlee2022autocrine,
  title={Autocrine signaling can explain the emergence of Allee effects in cancer cell populations},
  author={Gerlee, Philip and Altrock, Philipp M and Malik, Adam and Krona, Cecilia and Nelander, Sven},
  journal={PLoS computational biology},
  volume={18},
  number={3},
  pages={e1009844},
  year={2022},
  publisher={Public Library of Science San Francisco, CA USA}
}

@article{hanahan2000hallmarks,
  title={The hallmarks of cancer},
  author={Hanahan, Douglas and Weinberg, Robert A},
  journal={cell},
  volume={100},
  number={1},
  pages={57--70},
  year={2000},
  publisher={Elsevier}
}

@article{hoxhaj2020pi3k,
  title={The PI3K--AKT network at the interface of oncogenic signalling and cancer metabolism},
  author={Hoxhaj, Gerta and Manning, Brendan D},
  journal={Nature Reviews Cancer},
  volume={20},
  number={2},
  pages={74--88},
  year={2020},
  publisher={Nature Publishing Group UK London}
}

@article{hanahan2011hallmarks,
  title={Hallmarks of cancer: the next generation},
  author={Hanahan, Douglas and Weinberg, Robert A},
  journal={cell},
  volume={144},
  number={5},
  pages={646--674},
  year={2011},
  publisher={Elsevier}
}

@article{martin1997wound,
  title={Wound healing--aiming for perfect skin regeneration},
  author={Martin, Paul},
  journal={Science},
  volume={276},
  number={5309},
  pages={75--81},
  year={1997},
  publisher={American Association for the Advancement of Science}
}

@article{penzo2015organ,
  title={Organ-size regulation in mammals},
  author={Penzo-M{\'e}ndez, Alfredo I and Stanger, Ben Z},
  journal={Cold Spring Harbor perspectives in biology},
  volume={7},
  number={9},
  pages={a019240},
  year={2015},
  publisher={Cold Spring Harbor Lab}
}

@article{gokhale2015size,
  title={Size control: the developmental physiology of body and organ size regulation},
  author={Gokhale, Rewatee H and Shingleton, Alexander W},
  journal={Wiley Interdisciplinary Reviews: Developmental Biology},
  volume={4},
  number={4},
  pages={335--356},
  year={2015},
  publisher={Wiley Online Library}
}

@article{kimmel2026universal,
  title={Universal principles of cell population growth follow from local contact inhibition},
  author={Kimmel, Gregory J and Marzban, Sadegh and Damaghi, Mehdi and Traulsen, Arne and Anderson, Alexander RA and West, Jeffrey and Altrock, Philipp M},
  journal={iScience},
  volume={29},
  number={6},
  year={2026},
  publisher={Elsevier}
}

\newpage


\newpage
\renewcommand{\theequation}{S\arabic{equation}}
\renewcommand{\theHequation}{S\arabic{equation}}
\renewcommand{\theassumption}{SA\arabic{assumption}}
\renewcommand{\theHassumption}{S\arabic{assumption}}
\setcounter{figure}{0}
\setcounter{table}{0}
\setcounter{equation}{0}
\setcounter{assumption}{0}
\renewcommand{\figurename}{SI Fig.}

\phantomsection

\section*{Supplementary material for: Density-dependent growth emerges from Bayesian adaptation of
phenotype}
\begin{center}
    Manish Kumar Gupta, Arnab Barua and Haralampos Hatzikirou
\end{center}

\section*{S.1 Steady-state statistics of the bound complex and ligand}
\begin{table}[h]
\centering
\renewcommand{\arraystretch}{1.4}
\begin{tabular}{c l l}
\hline\hline
Symbol & Meaning & Units \\
\hline
$Y$        & Free (unbound) ligand count                          & count \\
$C_i$      & Bound complexes on cell $i$, $\le R_T$               & count \\
$R_T$      & Receptor count per cell                              & count \\
$N$        & Local population size (cell number)                  & count \\
$\alpha_Y$ & Maximal ligand production rate                       & count\,$\cdot$\,time$^{-1}$ \\
$d_Y$      & Ligand decay rate                                    & time$^{-1}$ \\
$ k_{\mathrm{on}}$ & Per-pair association rate                & time$^{-1}$ \\
$k_{\mathrm{off}}$     & Unbinding rate                           & time$^{-1}$ \\
$K_d$      & Half-occupancy ligand count, $k_{\mathrm{off}}/\hat k_{\mathrm{on}}$ & count \\
$K_N$      & Population size at half-maximal production            & count \\
\hline\hline
\end{tabular}
\caption{Symbols for the count-based ligand--receptor dynamics. All molecular
species are tracked as counts; the per-pair association rate
$\hat k_{\mathrm{on}}$ absorbs the reaction volume, so no volume appears, and the
half-occupancy ligand count is $K_d=k_{\mathrm{off}}/\hat k_{\mathrm{on}}$.}
\label{tab:ligand_symbols}
\end{table}

Each cell reads its environment through a diffusible ligand that binds to the
receptors on its surface; the bound complex is the signal the phenotype actually
responds to. What the phenotype sees is therefore not a single number but a
fluctuating count, set by the production of ligand, its capture and release at
the receptors, and the discreteness of these molecular events. In this section,
we derive the steady-state mean and variance of the free ligand $Y$ and the
bound-complex $C$, and track how both depend on the population size $N$. These
two moments are the input to the phenotypic dynamics developed later.

\subsection*{Molecular model}

Ligand is produced at a population-dependent rate, decays, and binds reversibly
to the $R_T$ receptors of each cell. Because the numbers of molecules are finite and each reaction occurs
stochastically, we track the state $(Y, C_1, \dots, C_N)$ with
the chemical Langevin equations
\begin{align}
dY_t &=
\Big[
\alpha_Y \tfrac{N}{N+K_N} - d_Y Y_t
- k_{\mathrm{on}} Y_t \!\sum_{i=1}^{N}(R_T - C_{i,t})
+ k_{\mathrm{off}} \!\sum_{i=1}^{N} C_{i,t}
\Big]dt \nonumber\\
&\qquad
+ \sqrt{
\alpha_Y \tfrac{N}{N+K_N} + d_Y Y_t
+ k_{\mathrm{on}} Y_t \!\sum_{i=1}^{N}(R_T - C_{i,t})
+ k_{\mathrm{off}} \!\sum_{i=1}^{N} C_{i,t}
}\; dW^{Y}_t ,
\label{eq:Y_sde}\\[6pt]
dC_{i,t} &=
\Big[
k_{\mathrm{on}} Y_t (R_T - C_{i,t}) - k_{\mathrm{off}} C_{i,t}
\Big]dt
+ \sqrt{
k_{\mathrm{on}} Y_t (R_T - C_{i,t}) + k_{\mathrm{off}} C_{i,t}
}\; dW^{C_i}_t .
\label{eq:Ci_sde}
\end{align}
Each elementary reaction event follows a Poisson process, so the variance of
each channel equals its rate, and independent channels add in quadrature; this
determines the noise amplitudes above. The complex equation is the same for every cell, and the cells are
coupled only through the ligand $Y_t$ they share.

The phenotype is the slow variable. It does not follow the ligand trajectory; it
responds to the signal once the fast chemistry has settled.
\begin{assumption}[Timescale separation]
\label{SIass:timescale}
The phenotype evolves on a timescale far longer than the relaxation timescales
of the ligand and the receptor--ligand binding dynamics,
\begin{equation}
(\tau_{\mathrm{bind}},\, \tau_Y) \;\ll\; \tau_{\mathrm{pheno}} .
\label{eq:timescales}
\end{equation}
\end{assumption}
\noindent
On the phenotypic timescale, $Y$ and $C$ have reached their stationary
distribution, so the phenotype reads their steady-state statistics. In the following, we obtain the steady state statistics of the bound-complex and ligand.

\subsection*{Linear noise approximation for obtaining steady-state statistics}

The mean of the ligand and bound-complex system follows from setting the drift terms of
Eqs.~\eqref{eq:Y_sde}--\eqref{eq:Ci_sde} to zero. To obtain the variance in the bound-complex, we linearize the full state $(Y,C_1,\dots,C_N)$ under van Kampen's system-size expansion, about the steady state

\begin{assumption}[Linear noise approximation]
\label{SIass:lna}
For large molecule numbers, the drift
is linearized about the steady-state means, and the noise amplitude is frozen at
that state. The deviations
\begin{equation}
\delta\mathbf{z}
= \big(\delta Y,\ \delta C_1,\ \dots,\ \delta C_N\big)^{\!\top},
\qquad
\delta C_i = C_i - \langle C\rangle_{\mathrm{ss}},
\end{equation}
are Gaussian and obey a linear equation with constant coefficients,
\begin{equation}
d\,\delta\mathbf{z} = \mathbf{J}\,\delta\mathbf{z}\,dt
+ \mathbf{B}\,d\mathbf{W}.
\label{eq:lna_sde}
\end{equation}
\end{assumption}
\noindent
The stationary covariance
$\boldsymbol{\Sigma}=\langle\delta\mathbf{z}\,\delta\mathbf{z}^{\!\top}\rangle$
solves the Lyapunov equation
\begin{equation}
\mathbf{J}\boldsymbol{\Sigma}
+ \boldsymbol{\Sigma}\mathbf{J}^{\!\top}
+ \mathbf{D} = 0, \qquad
\mathbf{D} = \mathbf{B}\mathbf{B}^{\!\top}.
\label{eq:lyapunov}
\end{equation}

To keep the $(N{+}1)\times(N{+}1)$ matrices readable, we collect the rates,
evaluated at the steady-state means, into
\begin{align}
\omega_0 &\equiv k_{\mathrm{on}}(R_T-\langle C\rangle_{\mathrm{ss}}), &
\omega_1 &\equiv k_{\mathrm{on}}\langle Y\rangle_{\mathrm{ss}}+k_{\mathrm{off}}, &
\omega_2 &\equiv d_Y, \nonumber\\
\omega_3 &\equiv k_{\mathrm{on}}\langle Y\rangle_{\mathrm{ss}}
(R_T-\langle C\rangle_{\mathrm{ss}})+k_{\mathrm{off}}\langle C\rangle_{\mathrm{ss}}, &
\omega_4 &\equiv 2\,d_Y\langle Y\rangle_{\mathrm{ss}}. &&
\label{eq:omega_defs}
\end{align}
Here $\omega_0$ is the per-cell association sensitivity, $\omega_1$ the
relaxation rate of one complex, $\omega_2$ the ligand decay rate, $\omega_3$ the
per-cell binding noise flux, and $\omega_4$ the ligand birth--death noise flux.

Linearizing the drift gives the relaxation matrix
\begin{equation}
\mathbf{J} =
\begin{pmatrix}
-(\omega_2+N\omega_0) & \omega_1 & \omega_1 & \cdots & \omega_1 \\
\omega_0 & -\omega_1 & 0 & \cdots & 0 \\
\omega_0 & 0 & -\omega_1 & \cdots & 0 \\
\vdots & \vdots & & \ddots & \vdots \\
\omega_0 & 0 & 0 & \cdots & -\omega_1
\end{pmatrix}.
\label{eq:J_explicit}
\end{equation}
Each cell binds the shared ligand, so $\omega_0$ sits in column~$0$ of every
complex row and the ligand diagonal collects all $N$ of them as
$-(\omega_2+N\omega_0)$. Each complex relaxes at $\omega_1$ on its own diagonal
and feeds back to the ligand through $\omega_1$ in row~$0$. The complexes do not
act on one another, so the lower-right block is diagonal.

The injected noise follows from the reaction stoichiometry: each channel carries
an independent white noise, so
$\mathbf{D}=\sum_r a_r\,\boldsymbol{\nu}_r\boldsymbol{\nu}_r^{\!\top}$ evaluated
at the means. Production and decay change only $Y$ and feed the $Y$--$Y$ entry,
$\omega_4$. Each binding or unbinding event has
$\boldsymbol{\nu}=\pm(\mathbf{e}_{C_i}-\mathbf{e}_Y)$ and adds $\omega_3$ to that
complex's diagonal, $\omega_3$ to the $Y$--$Y$ entry, and $-\omega_3$ to the
shared $Y$--$C_i$ entry. Distinct cells fire independent channels, so the
complexes share no noise. The diffusion matrix is
\begin{equation}
\mathbf{D} =
\begin{pmatrix}
\omega_4+N\omega_3 & -\omega_3 & -\omega_3 & \cdots & -\omega_3 \\
-\omega_3 & \omega_3 & 0 & \cdots & 0 \\
-\omega_3 & 0 & \omega_3 & \cdots & 0 \\
\vdots & \vdots & & \ddots & \vdots \\
-\omega_3 & 0 & 0 & \cdots & \omega_3
\end{pmatrix}.
\label{eq:D_explicit}
\end{equation}

Symmetry fixes the form of $\boldsymbol{\Sigma}$. The cells are identical and
enter $\mathbf{J}$ and $\mathbf{D}$ the same way, so $\mathrm{Var}(C_i)$ is the
same for every cell and $\mathrm{Cov}(C_i,C_j)$ the same for every pair. We
write these as $\mathrm{Var}(C)$ and $\mathrm{Cov}(C_i,C_j)$, and the covariance
reads
\begin{equation}
\boldsymbol{\Sigma} =
\begin{pmatrix}
\mathrm{Var}(Y) & \mathrm{Cov}(Y,C) & \mathrm{Cov}(Y,C) & \cdots & \mathrm{Cov}(Y,C) \\
\mathrm{Cov}(Y,C) & \mathrm{Var}(C) & \mathrm{Cov}(C_i,C_j) & \cdots & \mathrm{Cov}(C_i,C_j) \\
\mathrm{Cov}(Y,C) & \mathrm{Cov}(C_i,C_j) & \mathrm{Var}(C) & \cdots & \mathrm{Cov}(C_i,C_j) \\
\vdots & \vdots & \vdots & \ddots & \vdots \\
\mathrm{Cov}(Y,C) & \mathrm{Cov}(C_i,C_j) & \mathrm{Cov}(C_i,C_j) & \cdots & \mathrm{Var}(C)
\end{pmatrix}.
\label{eq:Sigma_ansatz}
\end{equation}
The four distinct entries reduce Eq.~\eqref{eq:lyapunov} to four scalar
equations for any $N$. Solving them gives
\begin{align}
\mathrm{Cov}(Y,C) &=
\frac{\omega_0\omega_4-\omega_2\omega_3}
{2\,\omega_2\,(\omega_1+\omega_2+N\omega_0)},
\label{eq:cov}\\[6pt]
\mathrm{Cov}(C_i,C_j) &=
\frac{\omega_0}{\omega_1}\,\mathrm{Cov}(Y,C),
\label{eq:covCC}\\[6pt]
\mathrm{Var}(C) &=
\frac{\omega_3}{2\,\omega_1}
+ \mathrm{Cov}(C_i,C_j),
\label{eq:varC}\\[6pt]
\mathrm{Var}(Y) &=
\frac{\omega_4+N\omega_3+2N\omega_1\,\mathrm{Cov}(Y,C)}
{2\,(\omega_2+N\omega_0)}.
\label{eq:varY}
\end{align}

Substituting rates in the original variables gives the stationary covariance
$\boldsymbol{\Sigma}$ with entries
\begin{align}
\mathrm{Cov}(Y,C) &=
\frac{k_{\mathrm{on}}\langle Y\rangle_{\mathrm{ss}}(R_T-\langle C\rangle_{\mathrm{ss}})
- k_{\mathrm{off}}\langle C\rangle_{\mathrm{ss}}}
{2\big[\,d_Y + N k_{\mathrm{on}}(R_T-\langle C\rangle_{\mathrm{ss}})
+ k_{\mathrm{on}}\langle Y\rangle_{\mathrm{ss}}+k_{\mathrm{off}}\,\big]} ,
\\[8pt]
\mathrm{Var}(C) &=
\frac{k_{\mathrm{on}}\langle Y\rangle_{\mathrm{ss}}(R_T-\langle C\rangle_{\mathrm{ss}})
+ k_{\mathrm{off}}\langle C\rangle_{\mathrm{ss}}}
{2\big(k_{\mathrm{on}}\langle Y\rangle_{\mathrm{ss}}+k_{\mathrm{off}}\big)}
+ \frac{k_{\mathrm{on}}(R_T-\langle C\rangle_{\mathrm{ss}})}
{k_{\mathrm{on}}\langle Y\rangle_{\mathrm{ss}}+k_{\mathrm{off}}}\,\mathrm{Cov}(Y,C) ,
\\[8pt]
\mathrm{Var}(Y) &=
\frac{d_Y\langle Y\rangle_{\mathrm{ss}}
+ \tfrac{N}{2}\big[k_{\mathrm{on}}\langle Y\rangle_{\mathrm{ss}}(R_T-\langle C\rangle_{\mathrm{ss}})
+ k_{\mathrm{off}}\langle C\rangle_{\mathrm{ss}}\big]
+ N\big(k_{\mathrm{on}}\langle Y\rangle_{\mathrm{ss}}+k_{\mathrm{off}}\big)\mathrm{Cov}(Y,C)}
{d_Y + N k_{\mathrm{on}}(R_T-\langle C\rangle_{\mathrm{ss}})} .
\end{align}
Within the linear-noise approximation, fluctuations are evaluated around the
deterministic steady state. The steady-state condition for the receptor--ligand
complex is
\begin{equation}
k_{\mathrm{on}}\langle Y\rangle_{\mathrm{ss}}
\bigl(R_T-\langle C\rangle_{\mathrm{ss}}\bigr)
-
k_{\mathrm{off}}\langle C\rangle_{\mathrm{ss}}
=0 .
\end{equation}
This is precisely the numerator appearing in the expression for
$\mathrm{Cov}(Y,C)$. Therefore,
\begin{equation}
\mathrm{Cov}(Y,C)=0 .
\end{equation}

\subsection*{Steady-state mean values}
Setting the deterministic part of the equation to zero,
we obtain the mean steady-state values
\begin{equation}
\boxed{\;
\langle Y\rangle_{\mathrm{ss}} = \frac{\alpha_Y}{d_Y}\,\frac{N}{N+K_N}.
\;}
\label{eq:Ymean}
\end{equation}
Thus, the population size enters the signal only through ligand production. The
bound-complex mean follows the Hill occupancy as a function of ligand abundance,
\begin{equation}
\boxed{\;
\langle C\rangle_{\mathrm{ss}}
= R_T\,\frac{\langle Y\rangle_{\mathrm{ss}}}{K_d+\langle Y\rangle_{\mathrm{ss}}} .
\;}
\label{eq:Cmean}
\end{equation}

\subsection*{Steady-state variance in the weak-binding limit}

Setting $\mathrm{Cov}(Y,C)=0$ in Eqs.~\eqref{eq:varY}--\eqref{eq:varC}, variances collapse to their bare values. The
free ligand keeps the Poisson statistics of its production and decay,
\begin{equation}
\mathrm{Var}(Y) = \langle Y\rangle_{\mathrm{ss}} ,
\label{eq:VarY_final}
\end{equation}
and the complex keeps the binomial variance of $R_T$ receptors each occupied with
probability $\langle C\rangle_{\mathrm{ss}}/R_T$,
\begin{equation}
\mathrm{Var}(C) = \langle C\rangle_{\mathrm{ss}}
\left(1-\frac{\langle C\rangle_{\mathrm{ss}}}{R_T}\right) .
\label{eq:VarC_binomial}
\end{equation}

\begin{assumption}[Weak binding limit]
    $K_d\gg \langle Y\rangle_{ss}$
\label{SIass:weak_binding}
\end{assumption}
Under the above assumption,  the occupancy is small, $\langle
C\rangle_{\mathrm{ss}}\ll R_T$, so the variance is well approximated as
\begin{equation}
\mathrm{Var}(C) \;\longrightarrow\; \langle C\rangle_{\mathrm{ss}} .
\;
\label{eq:VarC_final}
\end{equation}
At steady state, then, both the free ligand and the bound complex are Gaussian and
mutually uncorrelated, with means \eqref{eq:Ymean} and \eqref{eq:Cmean} carrying
all the dependence on population size.

\subsection*{Defining rescaled variables}
It is convenient to carry the free-ligand statistics in the notation of the
quasi-steady distribution and to measure the ligand in units of the
half-occupancy count $K_d$. Write the free-ligand mean and variance as
\begin{equation}
\mu_{Y,q} \equiv \langle Y\rangle_{\mathrm{ss}}
= \frac{\alpha_Y}{d_Y}\,\frac{N}{N+K_N},
\qquad
\sigma_{Y,q}^{2} \equiv \mathrm{Var}(Y) = \mu_{Y,q},
\label{eq:Y_moments_q}
\end{equation}
the variance equal to the mean by Poisson result \eqref{eq:VarY_final}.
Upon rescaling the ligand by $K_d$, we obtain
\begin{equation}
\tilde\mu_{Y,q} = \frac{\mu_{Y,q}}{K_d},
\qquad
\tilde\sigma_{Y,q}^{2} = \frac{\sigma_{Y,q}^{2}}{K_d^{2}}
= \frac{\tilde\mu_{Y,q}}{K_d}.
\label{eq:Y_moments_rescaled}
\end{equation}
Thus, the ligand is characterized by the dimensionless level $\tilde\mu_{Y,q}$ relative to half-occupancy.
The bound complex follows the Hill occupancy
in the rescaled ligand,
\begin{equation}
\langle  C\rangle
= R_T\,\frac{\tilde\mu_{Y,q}}{1+\tilde\mu_{Y,q}},
\qquad
\mathrm{Var}( C)
\approx \langle  C\rangle.
\label{eq:C_moments_rescaled}
\end{equation}

For notational convenience, the tilde is omitted in what follows, with all counts
expressed in $K_d$ units.

\newpage

\section*{S.2 A minimal decoder and the non-monotonic mismatch}

Section~S.1 established the signal read by a cell: the bound-complex count $C$,
which is approximately Gaussian-distributed and has a mean determined by
population size through the ligand level.
This section
asks what the cell does with that signal. A cell never observes $\mu_{Y,q}$ directly. Instead, it observes $C$ and must infer $\mu_{Y,q}$ by inverting the binding curve. The goal here is narrow and
specific: to show that the simplest standard decoder, applied to this signal,
produces a mismatch between the sensed and the true ligand that is
\emph{non-monotonic} in the signal level---large at low signal, small at
intermediate signal, large again at high signal. A single ingredient produces this behavior. A constant basal error in the sensed
bound-complex count generates this shape; removing it makes the mismatch
monotonic. We therefore state it
as an explicit assumption and motivate it biologically before deriving the
consequence.

\subsection*{Bound-complex statistics}

We carry over a single result from Section~S.1---the distribution of the bound
complex. At steady state, the mean bound-complex count follows the binding curve,
\begin{equation}
\langle C \rangle = R_T \frac{\mu_{Y,q}}{1 + \mu_{Y,q}},
\end{equation}
and its fluctuation is given by
\begin{equation}
\mathrm{Var}(C) = \langle C \rangle = R_T \frac{\mu_{Y,q}}{1 + \mu_{Y,q}}.
\end{equation}

\subsection*{Decoding bound-complex information into an extracellular ligand estimate}

The cell must turn the count $C$ into an estimate of the ligand $\mu_{Y,q}$. A standard choice in the receptor-sensing literature is to invert the
deterministic binding curve. Solving
$\langle C \rangle = R_T \mu_{Y,q}/(1+\mu_{Y,q})$ for the ligand gives the
estimator
\begin{equation}
\hat{\mu}_{Y,q} = f(C) = \frac{C}{R_T - C}.
\end{equation}
This is the natural inversion for a cell that reads receptor occupancy, and we
adopt it without modification.

\subsection*{Basal error}

A receptor is never silent. Nonspecific binding, ligand-independent receptor activation, and readout dark
counts all register as bound-complex counts, and the downstream decoder cannot
distinguish them from genuine binding. The cell therefore
inverts not just ligand-bound complex $C$ but a total count
\begin{equation}
C_{\text{total}} = C + E,
\end{equation}
where $E$ is a basal error. Each contributing event is discrete and
independent of true binding, so $E$ is itself Poisson, with mean and
variance both equal to $\epsilon$. It shifts the perceived mean and adds its own
spread,
\begin{align}
\langle C_{\text{total}} \rangle &= \langle C \rangle + \epsilon, \\
\mathrm{Var}(C_{\text{total}}) &= \mathrm{Var}(C) + \mathrm{Var}(\epsilon)
= \langle C \rangle + \epsilon.
\end{align}
This offset is the central mechanism underlying the non-monotonic dependence of
the mismatch derived below.

\subsection*{Estimation bias resulting from finite sampling and basal error}

The decoder is nonlinear, so a uncertain measurement of $C_{\text{total}}$
yields a biased estimate of $\mu_{Y,q}$. For $n$ independent reads, let $\bar{C}_{\text{total}}$ denote the sample mean.
We expand $f(\bar{C}_{\text{total}})$ to second order about
$\langle C_{\text{total}} \rangle$:
\begin{align}
f(\bar{C}_{\text{total}})
&\approx
f(\langle C_{\text{total}} \rangle)
+ f'(\langle C_{\text{total}} \rangle)
\bigl(\bar{C}_{\text{total}}-\langle C_{\text{total}} \rangle\bigr)
\nonumber\\
&\quad
+ \tfrac{1}{2}f''(\langle C_{\text{total}} \rangle)
\bigl(\bar{C}_{\text{total}}-\langle C_{\text{total}} \rangle\bigr)^2.
\end{align}
Upon averaging, the linear term vanishes, and the quadratic term contributes
through the variance of the sample mean:
$\mathrm{Var}(\bar{C}_{\text{total}}) = \mathrm{Var}(C_{\text{total}})/n$,
\begin{equation}
\mathbb{E}[\hat{\mu}_{Y,q}] \approx f(\langle C_{\text{total}} \rangle)
+ \tfrac{1}{2} f''(\langle C_{\text{total}} \rangle)\frac{\mathrm{Var}(C_{\text{total}})}{n}.
\end{equation}
Subtracting the true ligand splits the bias into a deterministic part and a
finite-sampling part,
\begin{equation}
\text{Bias}_{\text{total}} \approx
\underbrace{\left( \frac{\langle C \rangle + \epsilon}{R_T - (\langle C \rangle + \epsilon)} - \mu_{Y,q} \right)}_{\text{deterministic}}
+ \underbrace{\frac{R_T}{(R_T - \langle C \rangle - \epsilon)^3} \cdot \frac{\langle C \rangle + \epsilon}{n}}_{\text{finite sampling}}.
\end{equation}
We evaluate this expression in the regime used throughout the rest of the paper:
low ligand levels and a small basal background,
$\mu_{Y,q} \ll 1$ and $\epsilon \ll 1$. In this regime,
$\langle C \rangle \approx R_T\mu_{Y,q} \ll R_T$, and each denominator can be
approximated by $R_T$. The subleading corrections---a deterministic cross term $2\mu_{Y,q}\epsilon/R_T$
and a finite-sampling background term $\epsilon/nR_T^2$---are smaller than the
two retained terms by factors $2\mu_{Y,q}$ and $\epsilon/(\mu_{Y,q}R_T)$,
respectively, and we drop them. The bias reduces to
\begin{equation}
\text{Bias}_{\text{total}} \approx \frac{\epsilon}{R_T} + \frac{\mu_{Y,q}}{n R_T}.
\end{equation}
The first term is the background read as signal; it is constant in the ligand.
The second is the sampling bias; it grows with the ligand.

\subsection*{Estimation uncertainty due to finite sampling}

The uncertainty in the estimate follows from the same expansion at first order (the
delta method),
\begin{equation}
\mathrm{Var}(\hat{\mu}_{Y,q}) \approx
\left[ f'(\langle C_{\text{total}} \rangle) \right]^2 \mathrm{Var}(\bar{C}_{\text{total}}),
\qquad
f'(\langle C_{\text{total}}\rangle) = \frac{R_T}{(R_T - \langle C_{\text{total}}\rangle)^2}.
\end{equation}
In the same weak binding limit, $f' \to 1/R_T$. If the basal contribution to the
variance is negligible, then
$\mathrm{Var}(\bar{C}_{\text{total}}) \approx R_T\mu_{Y,q}/n$, so
\begin{equation}
\mathrm{Var}(\hat{\mu}_{Y,q}) \approx \frac{1}{R_T^2}\left( \frac{R_T \mu_{Y,q}}{n} \right)
= \frac{\mu_{Y,q}}{n R_T}.
\end{equation}

\subsection*{Baseline mismatch is a non-monotonic function of ligand level}

In the main text, we defined the baseline mismatch as the variance-normalized difference between the internally predicted mean signal at the reference phenotype $(X^*)$ and the actual extracellular mean signal,
\begin{equation}
\Delta_0
=
\frac{
\mu_{Y|X^*}-\mu_{Y,q}
}{
\sigma_Y
}.
\label{eq:baseline_mismatch_definition}
\end{equation}
Here, $\sigma_Y$ denotes the standard deviation of the marginal sensed-signal distribution entering the bivariate Gaussian closure. We now connect this effective quantity to the receptor-level sensing statistics derived above.

The receptor-level calculation gives the mean estimation bias at the reference phenotype,
\begin{equation}
\mu_{Y|X^*}-\mu_{Y,q}
\simeq
\frac{\epsilon}{R_T}
+
\frac{\mu_{Y,q}}{nR_T},
\label{eq:receptor_level_bias}
\end{equation}
together with the conditional sensing variance,
\begin{equation}
\sigma_{Y|X^*}^2
\simeq
\frac{\mu_{Y,q}}{nR_T}.
\label{eq:conditional_sensing_variance}
\end{equation}
For the bivariate Gaussian distribution used in the main text, the conditional and marginal variances are related by
\begin{equation}
\sigma_{Y|X^*}^2
=
\left(1-\rho^2\right)\sigma_Y^2.
\label{eq:conditional_marginal_variance_relation}
\end{equation}
Therefore,
\begin{equation}
\sigma_Y
=
\frac{\sigma_{Y|X^*}}{\sqrt{1-\rho^2}}
\simeq
\frac{1}{\sqrt{1-\rho^2}}
\sqrt{
\frac{\mu_{Y,q}}{nR_T}
}.
\label{eq:marginal_sensing_variance}
\end{equation}

Substituting Eqs.~\eqref{eq:receptor_level_bias} and
\eqref{eq:marginal_sensing_variance} into
Eq.~\eqref{eq:baseline_mismatch_definition} gives
\begin{align}
\Delta_0
&\simeq
\sqrt{1-\rho^2}
\frac{
\dfrac{\epsilon}{R_T}
+
\dfrac{\mu_{Y,q}}{nR_T}
}{
\sqrt{
\dfrac{\mu_{Y,q}}{nR_T}
}
}
\label{eq:baseline_mismatch_rho_corrected}
\end{align}

\paragraph{\textbf{Weak correlation regime.}} Throughout the main text, we work in the weak phenotype--signal correlation
regime, $\rho^2 \ll 1$. Consequently,
\begin{equation}
\sqrt{1-\rho^2}
=
1+\mathcal{O}(\rho^2)
\end{equation}
and the baseline mismatch reduces to
\begin{equation}
\Delta_0
\approx
\epsilon
\sqrt{
\frac{n}{\mu_{Y,q}R_T}
}
+
\sqrt{
\frac{\mu_{Y,q}}{nR_T}
}.
\label{eq:baseline_mismatch_weak_correlation}
\end{equation}

Two terms compete. The background term scales as $\mu_{Y,q}^{-1/2}$ and dominates
at low ligand; the sampling term scales as $\mu_{Y,q}^{1/2}$ and dominates at
high ligand. Differentiating,
\begin{equation}
\frac{d\Delta_0}{d\mu_{Y,q}} =
-\frac{1}{2}\sqrt{\frac{n}{R_T}}\frac{\epsilon}{\mu_{Y,q}^{3/2}}
+ \frac{1}{2}\sqrt{\frac{1}{n R_T}}\frac{1}{\mu_{Y,q}^{1/2}},
\end{equation}
and setting the derivative to zero locates the minimum,
\begin{equation}
\mu_{Y,q}^{\min} = n\epsilon, \qquad \Delta_{0}^{\min} = 2\sqrt{\frac{\epsilon}{R_T}}.
\end{equation}
 At low ligand the cell reads mostly background and
the mismatch is large. Raising the ligand buries the fixed background and the
mismatch falls, reaching its floor $2\sqrt{\epsilon/R_T}$ at
$\mu_{Y,q} = n\epsilon$. Past that point, finite-sampling shot noise sets the
accuracy, and the mismatch climbs again. The single basal error $\epsilon$ fixes
both the position and the depth of the optimum: send $\epsilon \to 0$ and the
low-ligand branch disappears, leaving the monotone sampling term alone.

\begin{figure}[pos=tbp!] \centering 
\includegraphics[width=0.9\linewidth]{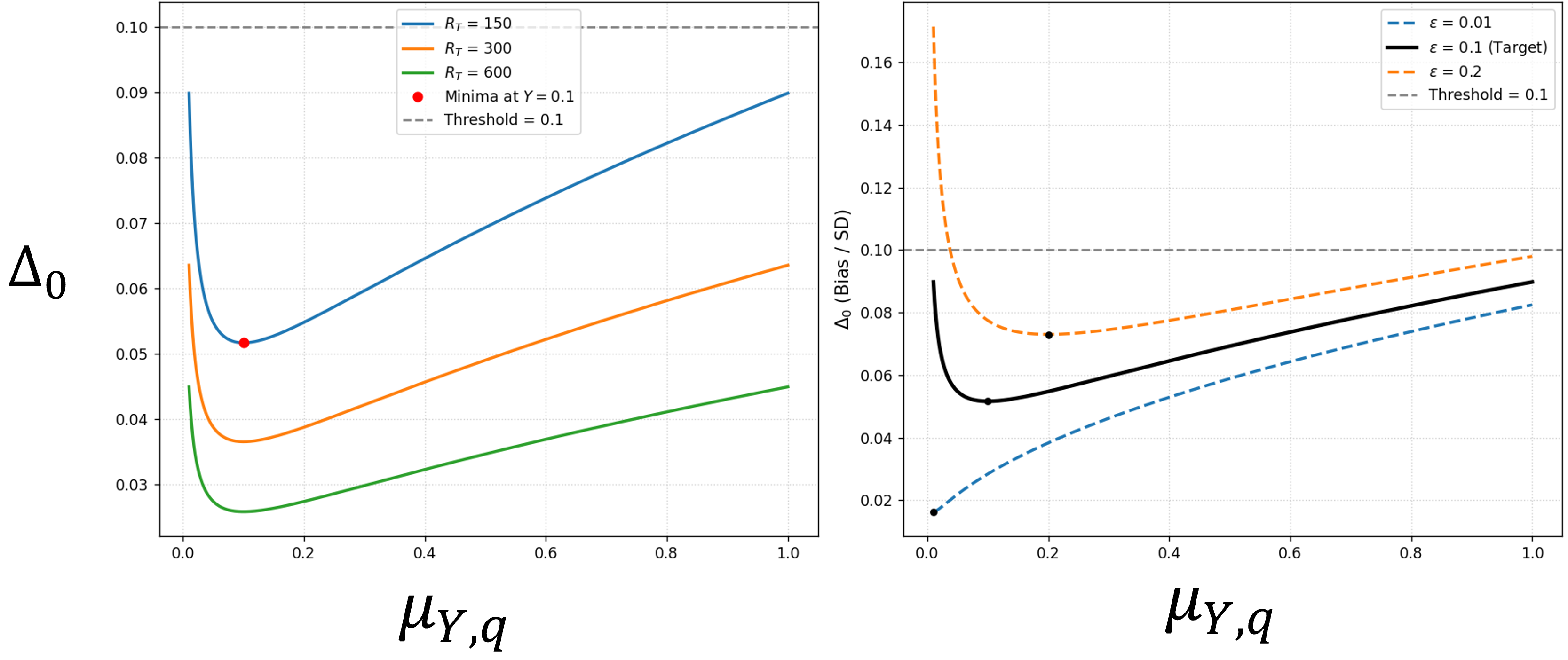} \caption{\textbf{Mismatch plot vs ligand level:}  Mismatch vs ligand level is plotted for varying receptor numbers (Left) and  mean basal error (right)} \label{SIfig:informationmismatch}
\end{figure}

\newpage

\section*{S.3 From single-cell adaptation to population-level phenotype dynamics}

At the single-cell level, the phenotype $X$ is a continuous internal state
that evolves through cell-autonomous adaptation. We write the probability
density of a single cell's phenotype as $p(X,t)$, normalized so that
$\int p(X,t)\,dX = 1$ at all times. Its evolution is governed by an
adaptation operator $\mathcal{L}$,
\begin{equation}
  \frac{\partial p(X,t)}{\partial t} = \mathcal{L}[p](X,t),
  \label{eq:single-cell}
\end{equation}
which encodes the deterministic drift and stochastic spread of the
phenotype as the cell senses its environment and reweights its internal
state. The operator $\mathcal{L}$ conserves probability:
$\int \mathcal{L}[p]\,dX = 0$, so Eq.~\eqref{eq:single-cell} preserves the
normalization of $p$.

We now ask how the \emph{population} phenotype distribution evolves once
cells also divide and die. Let $n(X,t)$ be the number density of cells in
phenotypic state $X$, so that the total population size is
\begin{equation}
  N(t) = \int n(X,t)\,dX,
\end{equation}
and define the population phenotype fraction
\begin{equation}
  \phi(X,t) = \frac{n(X,t)}{N(t)},
  \qquad \int \phi(X,t)\,dX = 1.
  \label{eq:phi-def}
\end{equation}
Crucially, $\phi$ is not the same object as $p$: the single-cell density
$p$ tracks one cell's internal adaptation, whereas $\phi$ is the phenotype
composition of the whole population, which is reshaped by proliferation
because cells in different states divide at different rates.

The number density obeys two processes. Each cell adapts according to the
same operator $\mathcal{L}$, and each cell in state $X$ contributes to net
growth at the per-capita rate $f(X)$, the net per-capita growth rate
(births minus deaths, which may be negative). Hence
\begin{equation}
  \frac{\partial n(X,t)}{\partial t}
  = \mathcal{L}[n](X,t) + f(X)\,n(X,t).
  \label{eq:n-evolution}
\end{equation}
The first term moves cells through phenotype space; the second changes
their number without moving them.

To obtain the dynamics of the composition $\phi$, we substitute
$n = \phi N$ into Eq.~\eqref{eq:n-evolution}. Since $\mathcal{L}$ acts only
on the phenotype argument and $N(t)$ is independent of $X$,
$\mathcal{L}[\phi N] = N\,\mathcal{L}[\phi]$, giving
\begin{equation}
  \frac{\partial (\phi N)}{\partial t}
  = N\,\mathcal{L}[\phi] + f(X)\,\phi N.
  \label{eq:phiN}
\end{equation}
Expanding the left-hand side with the product rule,
\begin{equation}
  \frac{\partial (\phi N)}{\partial t}
  = N\,\frac{\partial \phi}{\partial t} + \phi\,\frac{dN}{dt}.
  \label{eq:product-rule}
\end{equation}
The growth of $N$ follows from integrating Eq.~\eqref{eq:n-evolution} over
$X$. The adaptation term integrates to zero, leaving
\begin{equation}
  \frac{dN}{dt}
  = \int f(X)\,n(X,t)\,dX
  = N \int f(X)\,\phi(X,t)\,dX
  = N\,\bar{f},
  \label{eq:N-growth}
\end{equation}
where
\begin{equation}
  \bar{f}(t) = \int f(X)\,\phi(X,t)\,dX
  \label{seq:fbar}
\end{equation}
is the population-averaged net per-capita growth rate. Combining
Eqs.~\eqref{eq:phiN}--\eqref{eq:N-growth} and dividing through by $N$,
\begin{equation}
  \frac{\partial \phi}{\partial t}
  = \mathcal{L}[\phi]
  + f(X)\,\phi - \bar{f}\,\phi,
\end{equation}
which we write compactly as
\begin{equation}
  \boxed{\;
  \frac{\partial \phi(X,t)}{\partial t}
  = \mathcal{L}[\phi]
  + \bigl[\,f(X) - \bar{f}(t)\,\bigr]\,\phi(X,t).
  \;}
  \label{eq:phi-dynamics}
\end{equation}
Equation~\eqref{eq:phi-dynamics} is the central result of this subsection.
The adaptation operator $\mathcal{L}$ is inherited unchanged from the
single-cell dynamics, so individual plasticity enters the population
equation exactly as it acts on one cell. Proliferation enters through the
selection term $[f(X) - \bar{f}]\,\phi$: states that grow faster than the
population mean gain weight, states that grow slower lose it, and the
subtraction of $\bar{f}$ is precisely what keeps $\phi$ normalized. This is
the replicator structure of selection on a phenotypic landscape, and it
shows that population-level adaptation is not the single-cell operator
alone but $\mathcal{L}$ combined with growth-driven reweighting.

\newpage

\section*{S.4: Reducing the full phenotype-distribution dynamics to moment dynamics}
\label{si:moment_reduction}

The population phenotype density $\phi(X,t)$ obeys
\begin{equation}
\frac{\partial \phi}{\partial t}
=
\underbrace{
\frac{1}{\tau}
\left[\int q(Y;N)\frac{p(Y\mid X)}{p_{\mathrm{pop}}(Y,t)}\,dY - 1\right]\phi
}_{\text{Bayesian inference}}
+
\underbrace{
\gamma\,\partial_X\!\left[(X-X^*)\phi\right]
+D_X\,\partial_X^2\phi
}_{\text{intrinsic relaxation}}
+
\underbrace{
\left[f(X)-\bar{f}\right]\phi
}_{\text{natural selection}},
\label{si:full_phi}
\end{equation}
with $\bar{f}=\int f(X)\phi\,dX$ and
$p_{\mathrm{pop}}(Y,t)=\int p(Y\mid X)\phi(X,t)\,dX$.

Three processes act on $\phi$. The Bayesian-inference term reweights each
phenotype by how well it predicts the signal the population receives. The
intrinsic-fitness term relaxes the phenotype toward $X^*$ and spreads it by
diffusion. The natural-selection term multiplies each state by its growth rate
relative to the population mean. The first term is the obstacle. It integrates
the likelihood ratio $p(Y\mid X)/p_{\mathrm{pop}}$ against the environmental
signal distribution $q(Y;N)$, and $p_{\mathrm{pop}}$ is itself a functional of
$\phi$, so the equation does not close on its own. We reduce it to a closed pair
of equations for the phenotype mean $\mu_X$ and variance $\sigma_X^2$. Three assumptions lead to this reduction, and we introduce each one where it is needed in the derivation.

\subsection*{Gaussian closure and the small-coupling expansion}

The nonlinear terms in Eq.~\eqref{si:full_phi} generate a hierarchy in which the
dynamics of low-order moments depend on progressively higher-order moments. To
truncate this hierarchy, we assume that the phenotype distribution remains
sufficiently concentrated.

\begin{assumption}
\label{si:ass_gauss}
The phenotype density $\phi(X,t)$ is sufficiently concentrated that the joint
distribution of $(X,Y)$ can be approximated as bivariate Gaussian, with means
$(\mu_X,\mu_Y)$, variances $(\sigma_X^2,\sigma_Y^2)$, and correlation
coefficient $\rho$. In particular,
\begin{equation}
\left\langle (X-\mu_X)^3 \right\rangle = 0,
\qquad
\left\langle (X-\mu_X)^4 \right\rangle = 3\sigma_X^4.
\end{equation}
\end{assumption}

\noindent

Under the polynomial truncation introduced below, these Gaussian moment
relations close the hierarchy at the level of $\mu_X$ and $\sigma_X^2$. We
therefore define the standardized variables
\begin{equation}
z_X=\frac{X-\mu_X}{\sigma_X},
\qquad
z_Y=\frac{Y-\mu_Y}{\sigma_Y}.
\end{equation}

The Bayesian-inference term nevertheless retains the full likelihood ratio. We
simplify this ratio by expanding it in the strength of the phenotype--signal
coupling.

\begin{assumption}
\label{si:ass_coupling}
The phenotype and signal are weakly coupled, such that $\rho^2\ll 1$. The
likelihood ratio may then be expanded as
\begin{equation}
\frac{p(Y\mid X)}{p_{\mathrm{pop}}(Y,t)}
=
1+\rho\,z_Xz_Y
+\frac{\rho^2}{2}(z_X^2-1)(z_Y^2-1)
+\mathcal{O}(\rho^3).
\label{si:LR}
\end{equation}
\end{assumption}

\noindent
This expansion is ordered in powers of $\rho$. The leading correction is
bilinear in the standardized variables, whereas the second-order term couples
their second Hermite factors. Thus, $\rho$ serves as the small parameter
controlling the reduction.

\subsection*{Averaging over the environmental signal}

The likelihood ratio in Eq.~\eqref{si:LR} is averaged over the environmental
signal distribution $q(Y;N)$ introduced in Section~S.1. At population size $N$,
the received signal is Gaussian, with mean $\mu_{Y,q}$ and variance
\begin{equation}
\sigma_{Y,q}^2=\frac{\mu_{Y,q}}{K_d}.
\end{equation}
This distribution characterizes the signal received by the population. By
contrast, the predictive distribution $p_{\mathrm{pop}}(Y,t)$ has mean $\mu_Y$
and standard deviation $\sigma_Y$. These quantities need not coincide with the
corresponding moments of $q(Y;N)$. We quantify the discrepancy between the
predicted and received signal means by the standardized mismatch
\begin{equation}
\Delta
:=
\frac{\mu_Y-\mu_{Y,q}}{\sigma_Y}.
\label{si:Delta_def}
\end{equation}
Hence, $\Delta>0$ indicates that the population-level predictive distribution
overestimates the mean received signal.

Averaging Eq.~\eqref{si:LR} over $q(Y;N)$ replaces the dependence on $z_Y$ by
its moments under the environmental signal distribution. Two additional
smallness conditions simplify this average.

\begin{assumption}
\label{si:ass_fluct}
Environmental fluctuations are small relative to the predictive spread,
$\sigma_{Y,q}^2\ll\sigma_Y^2$, and the standardized mismatch is small,
$\Delta^2\ll 1$. Under these conditions,
\begin{equation}
\left\langle z_Y \right\rangle_q \simeq -\Delta,
\qquad
\left\langle z_Y^2 \right\rangle_q \simeq \Delta^2.
\end{equation}
Neglecting terms of order $\Delta^2$ then gives
\begin{equation}
\int q(Y;N)
\left[
\frac{p(Y\mid X)}{p_{\mathrm{pop}}(Y,t)}-1
\right]dY
=
\underbrace{-\rho\Delta\,z_X}_{\mathcal{O}(\rho)}
\underbrace{-\frac{\rho^2}{2}(z_X^2-1)}_{\mathcal{O}(\rho^2)}
+
\mathcal{O}(\rho^3,\rho^2\Delta^2).
\label{si:avg_LR}
\end{equation}
\end{assumption}

\noindent
Two contributions remain at the retained order. The term of order
$\mathcal{O}(\rho)$ is linear and odd in $z_X$, and its magnitude is controlled
by the mismatch $\Delta$. This term contributes to the dynamics of the mean
phenotype. The term of order $\mathcal{O}(\rho^2)$ is even in $z_X$ and
independent of $\Delta$ at leading order. It contributes to the dynamics of the
phenotypic variance.

\subsection*{Reduced phenotype-distribution equation}

Substituting Eq.~\eqref{si:avg_LR} into Eq.~\eqref{si:full_phi} and adopting the
quadratic fitness landscape
\begin{equation}
f(X)=f_0-\alpha(X-X^*)^2
\end{equation}
gives
\begin{equation}
\frac{\partial\phi}{\partial t}
=
\underbrace{
\frac{1}{\tau}
\left[
-\rho\Delta\,z_X
-\frac{\rho^2}{2}(z_X^2-1)
\right]\phi
}_{\text{Bayesian inference}}
+
\underbrace{
\gamma\,\partial_X\!\left[(X-X^*)\phi\right]
+D_X\partial_X^2\phi
}_{\text{intrinsic relaxation}}
+
\underbrace{
\left[
f_0-\alpha(X-X^*)^2-\bar{f}
\right]\phi
}_{\text{natural selection}}.
\label{si:reduced_phi}
\end{equation}

Equation~\eqref{si:reduced_phi} is local in $\phi$ and polynomial in $z_X$.
Under Assumption~\ref{si:ass_gauss}, its first two moments therefore form a
closed dynamical system.

\subsection*{Moment dynamics}

Assuming that $\phi$ is normalized, the mean phenotype is
\begin{equation}
\mu_X=\int X\phi(X,t)\,dX.
\end{equation}
Multiplying Eq.~\eqref{si:reduced_phi} by $X$ and integrating over phenotype
space gives
\begin{equation}
\dot{\mu}_X
=
\int X\,\frac{\partial\phi}{\partial t}\,dX.
\end{equation}
Evaluating each contribution yields
\begin{equation}
\dot{\mu}_X
=
\underbrace{
-\frac{\rho\Delta}{\tau}\sigma_X
}_{\text{Bayesian inference}}
\underbrace{
-\gamma(\mu_X-X^*)
}_{\text{intrinsic relaxation}}
\underbrace{
-2\alpha(\mu_X-X^*)\sigma_X^2
}_{\text{natural selection}}.
\label{si:mudot}
\end{equation}
The Bayesian contribution of order $\mathcal{O}(\rho^2)$ vanishes because it is
proportional to
\begin{equation}
\left\langle z_X(z_X^2-1)\right\rangle=0,
\end{equation}
which follows from the Gaussian closure in
Assumption~\ref{si:ass_gauss}.

The phenotypic variance is
\begin{equation}
\sigma_X^2
=
\int (X-\mu_X)^2\phi(X,t)\,dX.
\end{equation}
Its time derivative may be written as
\begin{equation}
\dot{\sigma}_X^2
=
\int (X-\mu_X)^2\frac{\partial\phi}{\partial t}\,dX,
\end{equation}
because the additional term arising from the time dependence of $\mu_X$
vanishes by
$\int (X-\mu_X)\phi\,dX=0$. Evaluating the three contributions gives
\begin{equation}
\dot{\sigma}_X^2
=
\underbrace{
-\frac{\rho^2}{\tau}\sigma_X^2
}_{\text{Bayesian inference}}
\underbrace{
-2\gamma\sigma_X^2+2D_X
}_{\text{intrinsic relaxation}}
\underbrace{
-2\alpha\sigma_X^4
}_{\text{natural selection}}.
\label{si:sigdot}
\end{equation}
The Bayesian contribution of order $\mathcal{O}(\rho)$ vanishes by symmetry,
whereas the contribution of order $\mathcal{O}(\rho^2)$ follows from
\begin{equation}
\left\langle z_X^2(z_X^2-1)\right\rangle
=
\left\langle z_X^4\right\rangle
-
\left\langle z_X^2\right\rangle
=
3-1
=
2.
\end{equation}

Equations~\eqref{si:mudot} and \eqref{si:sigdot} therefore provide a closed
description of the phenotype-distribution dynamics at the level of the mean
and variance, valid to second order in the weak phenotype--signal coupling
$\rho$ and to leading order in the standardized mismatch $\Delta$.

\subsection*{Validating the approximate moment dynamics against the full PDE dynamics}

To validate the reduced moment equations, we compare their solutions with a
direct numerical solution of the full phenotype-distribution dynamics in
Eq.~\eqref{si:full_phi}. To simulate the full PDE, we must specify the
conditional sensing distribution $p(Y\mid X)$. For simplicity, we neglect the
variance of the received environmental signal and approximate
\begin{equation}
q(Y;N)
\simeq
\delta\!\left(Y-\mu_{Y,q}(N)\right).
\label{si:q_delta_validation}
\end{equation}
The Bayesian-inference term in Eq.~\eqref{si:full_phi} then becomes
\begin{equation}
\int q(Y;N)
\left[
\frac{p(Y\mid X)}{p_{\mathrm{pop}}(Y,t)}
-1
\right]dY
=
\frac{p(\mu_{Y,q}\mid X)}
     {p_{\mathrm{pop}}(\mu_{Y,q},t)}
-1.
\label{si:bayesian_validation}
\end{equation}

Following the linear Gaussian measurement model introduced in the main text,
the conditional sensing distribution evaluated at the received signal is
\begin{equation}
p(\mu_{Y,q}\mid X)
=
\frac{1}{\sqrt{2\pi}\sigma_Y}
\exp\!\left[
-\frac{1}{2}
\left(
\frac{\mu_{Y,q}-\mu_{Y\mid X}}{\sigma_Y}
\right)^2
\right].
\label{si:conditional_kernel_validation}
\end{equation}
The standardized difference appearing in this expression is
\begin{equation}
\frac{\mu_{Y,q}-\mu_{Y\mid X}}{\sigma_Y}
=
-\Delta(X).
\label{si:Delta_X_validation}
\end{equation}
As shown in the main text, the linear measurement model gives
\begin{equation}
\Delta(X)
=
\Delta_0
+
\rho\frac{X-X^*}{\sigma_X},
\label{si:Delta_X_Delta0_validation}
\end{equation}
where
\begin{equation}
\Delta_0
=
\frac{\mu_{Y\mid X^*}-\mu_{Y,q}(N)}{\sigma_Y}
\label{si:Delta0_validation}
\end{equation}
is the mismatch evaluated at the intrinsic phenotype $X^*$. Therefore,
Eq.~\eqref{si:conditional_kernel_validation} can be written as
\begin{equation}
p(\mu_{Y,q}\mid X)
=
\frac{1}{\sqrt{2\pi}\sigma_Y}
\exp\!\left[
-\frac{1}{2}
\left(
\Delta_0(N)
+
\rho\frac{X-X^*}{\sigma_X}
\right)^2
\right].
\label{si:conditional_kernel_Delta0_validation}
\end{equation}
At each time step, the population-level predictive density entering the full
PDE is evaluated self-consistently as
\begin{equation}
p_{\mathrm{pop}}(\mu_{Y,q},t)
=
\int
p(\mu_{Y,q}\mid X)\phi(X,t)\,dX.
\label{si:ppop_validation}
\end{equation}

The mismatch entering the reduced moment equations is obtained by averaging
Eq.~\eqref{si:Delta_X_Delta0_validation} over the phenotype distribution:
\begin{equation}
\Delta(\mu_X)
=
\Delta_0(N)
+
\rho\frac{\mu_X-X^*}{\sigma_X}.
\label{si:Delta_moment_validation}
\end{equation}
Thus, the same baseline mismatch $\Delta_0(N)$ determines the conditional
sensing distribution used in the full PDE and the effective mismatch entering
the reduced moment equations. We solve the full PDE and
Eqs.~\eqref{si:mudot}--\eqref{si:sigdot} from the same initial phenotype
distribution and compare the resulting dynamics of $\mu_X(t)$ and
$\sigma_X^2(t)$.

\begin{figure}[pos=tbp!]
\centering
\includegraphics[width=0.9\linewidth]{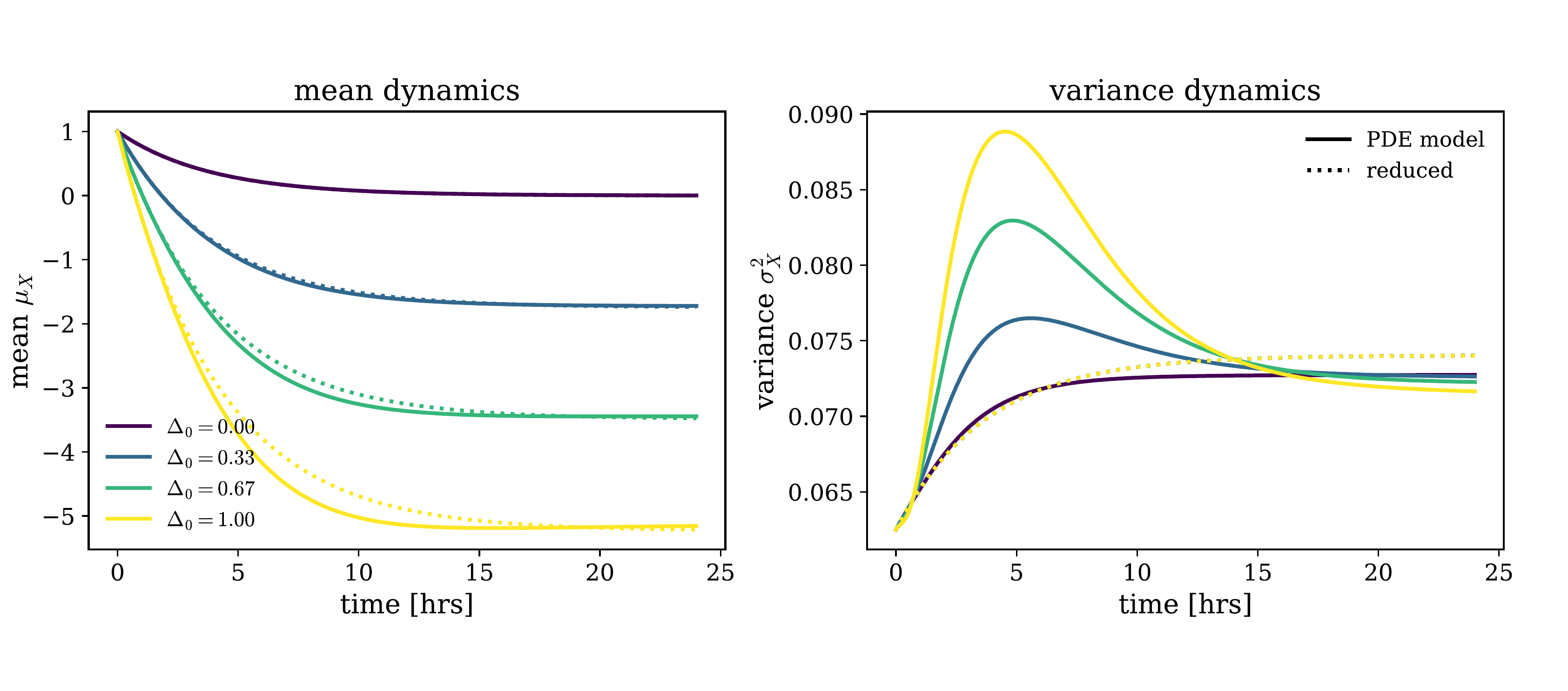}
\caption{\textbf{Reduced moment dynamics reproduce the full phenotype-density
dynamics.} Numerical solutions of the full PDE \eqref{si:full_phi} (solid lines) are
compared with solutions of the reduced moment equations
\eqref{si:mudot}--\eqref{si:sigdot} (dashed lines). \textbf{Left:} mean phenotype $\mu_X(t)$. \textbf{Right:} phenotype
variance $\sigma_X^2(t)$. Both panels share the same simulation, started from a
Gaussian initial density. Parameters: intrinsic relaxation rate $\gamma=0.05$ and
phenotypic diffusion $D_X=0.05$, with phenotype--signal coupling $\rho=0.005$,
mismatch, inference timescale $\tau=0.02$, selection strength
$\alpha=0.001$. The two descriptions agree
closely in the small-mismatch regime $\Delta_0^2\ll1$; the deviation grows with
$\Delta_0$, consistent with the small-$\Delta_0$ assumption used in the reduction.}
\label{SIfig:reducedynamicsfigure}
\end{figure}

SI Fig.~\ref{SIfig:reducedynamicsfigure} compares numerical solutions of the full PDE
\eqref{si:full_phi} with solutions of the reduced moment equations
\eqref{si:mudot}--\eqref{si:sigdot}, tracking both $\mu_X(t)$ and $\sigma_X^2(t)$.
The two agree well in the small-mismatch regime $\Delta_0^2\ll 1$; for larger $\Delta_0$
the moment equations deviate slightly, consistent with the small $\Delta_0$ assumption.

\newpage

\subsection*{S.5: Cellular Sensing and Phenotypic Adaptation via Minimal Regulatory Update}

We propose that single-cell adaptation follows a minimal regulatory update principle: over a finite observation window, the cell changes its effective occupancy over phenotypic states only as much as necessary for the signal statistics implied by its current phenotypic organization to match the signal statistics encountered in the environment. 
In this view, adaptation is conservative rather than arbitrary: once the cell's perceived signal distribution agrees with the empirical environmental signal distribution, no further update is required.

To formalize this idea, we consider a \emph{single cell} with latent phenotypic state \(X\) and external signal \(Y\), where \(Y\) is registered through receptor--ligand interactions and downstream signaling. The quantity \(p(x,t)\) denotes the cell's effective occupancy distribution over phenotypic states on the timescale of sensing and adaptation. Biologically, this distribution may be interpreted as a coarse-grained occupancy over metastable phenotypic programs accessible to the cell during the observation interval.

For definiteness, conditioned on phenotypic state \(X\), the receptor-level sensing of signal is assumed to follow a noisy linear relation
\begin{equation}
    Y = mX + c + \xi,
\end{equation}
where \(m\) is the coupling between phenotype and sensed signal, \(c\) is a baseline offset, and \(\xi\) represents stochastic fluctuations arising from receptor binding and downstream signal transduction. This induces the internal generative model
\begin{equation}
    p(X,Y)=p(X)\,p(Y\mid X),
\end{equation}
which specifies how the cell's current phenotypic organization predicts receptor-level signal statistics.

The corresponding \emph{perceived signal distribution} is
\begin{equation}
    p(Y)=\int p(X,Y)\,dx.
\end{equation}
This is the distribution of signals predicted by the cell's current phenotypic organization. Over the same observation window, however, the environment generates an \emph{empirical environmental signal distribution}, denoted by \(q(Y)\). In general, the perceived signal distribution \(p(Y)\) need not coincide with the empirical environmental signal distribution \(q(Y)\).

We assume that adaptation acts by reweighting the occupancy \(p(X)\) while holding the observation model \(p(Y\mid X)\) fixed. Thus, the cell updates its joint model \(p(X,Y)\) to a new distribution \(p_{\mathrm{new}}(X,Y)\) such that the perceived signal statistics become consistent with the empirically encountered environmental statistics.

\paragraph{\textbf{Variational formulation.}}

We represent this principle as a minimum-information update:
\begin{equation}
    p_{\mathrm{new}}(X,Y)
    =
    \arg\min_{p'} D_{\mathrm{KL}}\!\left(p'(X,Y)\,\|\,p(X,Y)\right),
\end{equation}
subject to the marginal constraint
\begin{equation}
    \int p'(X,Y)\,dx = q(Y), \qquad \forall y,
\end{equation}
together with normalization of \(p'(X,Y)\). We assume throughout that \(q\) is absolutely continuous with respect to \(p\) on the support of \(Y\), so that \(q(Y)/p(Y)\) is well defined whenever \(q(Y)>0\).

This variational problem expresses the idea that adaptation changes the cell's effective occupancy over phenotypic states only as much as needed to make the model-implied signal statistics agree with the empirically encountered environmental statistics.

\paragraph{\textbf{Solution of the constrained problem.}}

Introducing a Lagrange multiplier \(\lambda(y)\) for the marginal constraint and a scalar multiplier \(\alpha\) for normalization, the stationary solution has the form
\begin{equation}
    p_{\mathrm{new}}(X,Y)=p(X,Y)\,C(Y),
\end{equation}
where \(C(Y)=\exp(\lambda(Y)+\alpha-1)\) depends only on \(Y\). Enforcing the marginal constraint gives
\begin{equation}
    C(Y)=\frac{q(Y)}{p(Y)}.
\end{equation}
Hence,
\begin{equation}
\boxed{
    p_{\mathrm{new}}(X,Y)
    =
    p(X,Y)\,\frac{q(Y)}{p(Y)}
    =
    q(Y)\,p(X\mid Y).
}
\label{eq:updated_joint}
\end{equation}

Thus, the empirical environmental signal distribution \(q(Y)\) replaces the cell's perceived signal distribution \(p(Y)\), while the conditional structure \(p(X\mid Y)\) inherited from the original model is preserved.

\paragraph{\textbf{Updated occupancy over phenotypic states.}}

Marginalizing Eq.~\eqref{eq:updated_joint} over \(Y\) yields
\begin{equation}
\boxed{
    p_{\mathrm{new}}(X)
    =
    \int q(Y)\,p(X\mid Y)\,dy
    =
    p(X)\int q(Y)\,\frac{p(Y\mid X)}{p(Y)}\,dY.
}
\label{eq:updated_marginal}
\end{equation}
This expression makes the biological interpretation transparent: the cell does not alter the empirical environmental signal statistics \(q(Y)\); instead, it reweights phenotypic states according to how compatible their predicted signal responses are with the signals actually encountered.

If \(q(Y)=\delta(Y-Y_0)\), then Eq.~\eqref{eq:updated_marginal} reduces to
\begin{equation}
    p_{\mathrm{new}}(X)=p(X\mid Y_0),
\end{equation}
which is ordinary Bayesian conditioning on a single observed signal. For a general distribution \(q(Y)\), the updated occupancy is a weighted average of posterior phenotypic states across the range of signals encountered during the observation window.

\paragraph{\textbf{Sequential update in discrete time.}}

Let \(p(X,t)\) denote the effective occupancy distribution over phenotypic states at time \(t\). Over the interval \([t,t+\tau]\), the environment generates an empirical environmental signal distribution \(q_t(Y)\), while the cell's current phenotypic organization implies the perceived signal distribution
\begin{equation}
    p_t(Y)=\int p(X,t)\,p(Y\mid X)\,dX.
\end{equation}
Applying Eq.~\eqref{eq:updated_marginal} iteratively gives
\begin{equation}
    p(X,t+\tau)
    =
    p(X,t)\,\Lambda_t(X),
\label{eq:discrete_update}
\end{equation}
where
\begin{equation}
    \Lambda_t(X)
    :=
    \int q_t(Y)\,\frac{p(Y\mid X)}{p_t(Y)}\,dY.
\end{equation}
The factor \(\Lambda_t(x)\) increases the weight of phenotypic states whose predicted signal responses are more compatible with the empirical environmental signal distribution and decreases the weight of less compatible states. Normalization is preserved automatically:
\begin{equation}
    \int p(X,t+\tau)\,dX = 1.
\end{equation}

If the perceived signal distribution already matches the empirical environmental signal distribution, that is,
\begin{equation}
    q_t(Y)=p_t(Y),
\end{equation}
then
\begin{equation}
    \Lambda_t(X)=1
\end{equation}
for all \(X\), and no update occurs. Adaptation therefore stops once the signal statistics implied by the cell's current phenotypic organization are consistent with those generated by the environment.

\newpage

\begin{table}[ht]
\caption{Model parameters used in the simulations. Unless otherwise stated, all parameters are held fixed at the values listed. Parameters marked with $\dagger$ are varied in the results section to analyse their effect on growth regulation.}
\centering
\label{tab:model_parameters}
\begin{tabular}{c l l l}
\hline
\textbf{Symbol} & \textbf{Description} & \textbf{Value} & \textbf{Unit} \\
\hline
$f_0$                          & Bare proliferation rate                          & $0.002$    & $\mathrm{h^{-1}}$ \\
$\alpha$                       & Fitness curvature                                & $0.001$   & $\mathrm{h^{-1}}$ \\
$\gamma$                       & Intracellular regulatory relaxation rate          & $10^{-2}$ & $\mathrm{h^{-1}}$ \\
$D_X$                          & Phenotype diffusion coefficient                  & $10^{-2}$ & $\mathrm{h^{-1}}$ \\
$\tau$                          & Bayesian time scale                & 0.02 & $\mathrm{h}$ \\
$\rho^{\dagger}$               & Phenotype--signal correlation                     & $0.02$    & -- \\
$R_T$                          & Total receptor number per cell                   & $200$     & -- \\
$n$          & Number of independent receptor samples           & $1$      & -- \\
$\epsilon^{\dagger}$           & Basal readout offset                             & $0.07$    & -- \\
$Y_{\max}$                     & Maximal rescaled ligand concentration            & $0.3$     & -- \\
$K_N$                          & Half-maximal production population size          & $1000$    & -- \\
\hline
\end{tabular}
\end{table}

\newpage

\end{document}